%% file: main.tex
\pdfoutput=1
\RequirePackage{fix-cm}
\documentclass[twocolumn]{svjour3}

\smartqed  % flush right qed marks, e.g. at end of proof

\usepackage{wrapfig}
\usepackage{graphicx}
\usepackage{enumitem}

\usepackage{algorithmicx}
\usepackage{algorithm}
\usepackage{algpseudocode}

\usepackage{comment}

\usepackage{amsmath}
\usepackage{color}
\usepackage{xcolor}
\usepackage{url}
\usepackage{multirow}
\usepackage{subcaption}
\usepackage{hhline}
\usepackage{balance}
\usepackage{colortbl}
\usepackage{units}
\usepackage{marginnote}
\usepackage{pgfplotstable,array}
\usepackage{marginnote}
\usepackage{float}
\usepackage{pgfplots}
\usepackage{pgf,tikz}
\usepackage{mathtools}
\usepackage{xstring}
\usetikzlibrary{spy}
\usetikzlibrary{patterns}

\newcommand{\rev}[1]{#1}

\newcommand{\revj}[1]{#1}

\newcommand{\oldmethod}{RI}
\newcommand{\methodname}{APRIL}

\newcommand{\hide}[1]{}

\newcommand{\stitle}[1]{\noindent\textup{\textbf{#1}}}

\makeatletter
\newcommand{\currentfsize}{\f@size pt}
\makeatother

\newdimen\fsize

\graphicspath{ {./figures/} }

\begin{document}

\title{Raster Interval Object Approximations for Spatial Intersection Joins}

\author{Thanasis Georgiadis \and Eleni {Tzirita Zacharatou} \and Nikos Mamoulis}

\institute{T. Georgiadis \at
  Department of Computer Science \&
  Engineering, University of Ioannina, Greece\\
  \email{ageorgiadis@cs.uoi.gr}
  \and E. {Tzirita Zacharatou} \at
  IT University of Copenhagen, Denmark\\
 \email{elza@itu.dk}
\and
           N. Mamoulis \at
           Department of Computer Science \& Engineering,
           University of Ioannina, Greece\\
           \email{nikos@cs.uoi.gr}}  %\\

\date{Received: date / Accepted: date}

\maketitle

\begin{abstract}
Spatial join processing techniques that identify intersections between
complex geometries (e.g., polygons) commonly follow a two-step
filter-and-refine
\linebreak
pipeline.
The filter step evaluates the query predicate on the minimum
bounding rectangles (MBRs) of the geometries, while the refinement
step eliminates false positives by applying the query on the exact geometries.
\revj{To accelerate spatial join evaluation over complex geometries, we propose a {\em raster intervals approximation}} of object
geometries and introduce  a powerful {\em intermediate} step
in the pipeline.
In a preprocessing phase, our method (i) rasterizes each object
geometry using a fine grid, (ii) models groups of nearby cells that
intersect the polygon as an interval, and (iii) encodes each interval
with a bitstring capturing the overlap of each cell in it with the
polygon.
Going one step further, we improve our approach by approximating each object
by two sets of intervals that succinctly capture the raster cells which
(i) intersect with the object and (ii) are fully contained within the
object.
Using this representation, we show that we can verify whether two
polygons intersect through a sequence of linear-time joins between the
interval sets.
Our approximations are effectively compressible and customizable for partitioned data and polygons of varying sizes, 
rasterized at different granularities.
Finally, we propose a novel algorithm that
computes the interval approximation of a polygon
without fully rasterizing it first, rendering the computation of
approximations orders of magnitude faster.
Experiments on real data demonstrate the effectiveness
and efficiency of our proposal over previous work.
\end{abstract}

\input{introduction}
\input{background}

\input{ri}

\input{methodology}

\input{customization}
\input{rasterization}
\input{experiments}
\input{relatedwork}
\input{conclusion}

\vspace{4mm}
\noindent
\includegraphics[width=0.08\textwidth]{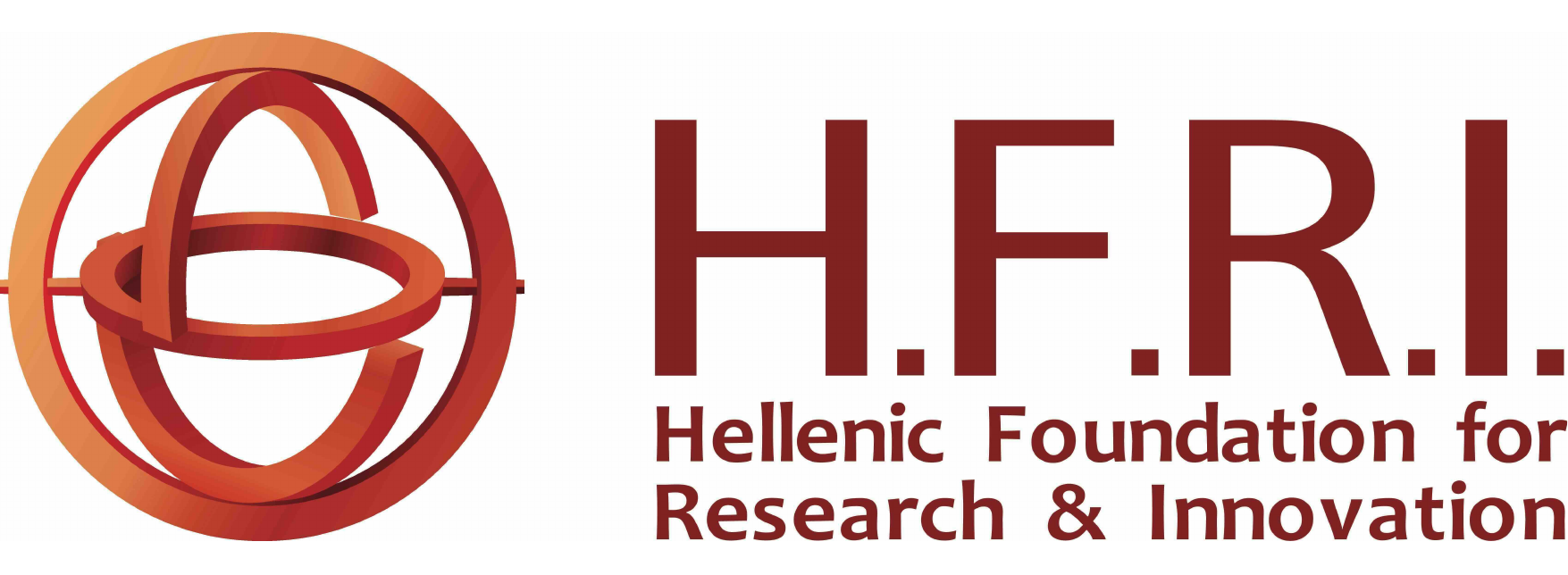}
\vspace{-4mm}
\begin{acknowledgements}
The research project was supported by the Hellenic Foundation
for Research and Innovation\linebreak (H.F.R.I.) under the ``2nd Call for
H.F.R.I. Research Projects to support Faculty Members \&
Researchers'' (Project Number: 02757).
\end{acknowledgements}

%% the bibliography file.
\bibliographystyle{spmpsci}      % mathematics and physical sciences
\bibliography{references}

\end{document}

%% file: introduction.tex
\section{Introduction}\label{sec:intro}
We study the problem of computing the spatial intersection join
between two spatial object collections $R$ and $S$, which identifies all pairs of
objects $(r,s), r\in R, s\in S$ such that $r$ shares at least one
common point with $s$.
Besides being a common operation in geographic information systems
(GIS),
the spatial intersection join
finds a wide range of applications in geo-spatial interlinking \cite{Papadakis21},
GeoSPARQL queries on RDF data stores \cite{TheocharidisLMB19},
interference detection between objects in computer graphics
\cite{ShinyaF91}, and the suggestion of synapses between neurons in
neuroscience models \cite{NobariTHKBA13}.
Recently, there has been a growing interest in spatial query evaluation over
complex object geometries, such as polygons
\cite{DoraiswamyF20,DoraiswamyF22,KipfLPPAZDB0K20,Kipf21,NobariQJ17,PandeyKNK18,SidlauskasCZA18,Teng21,ZacharatouDASF17,ZacharatouKSPDM21}.

A naive way to evaluate the join is to run an intersection test
algorithm from computational geometry for each pair $(r,s)$ in
$R\times S$. However, this method is extremely expensive,
since (i) the
number $|R\times S|$ of pairs to be tested can be huge and (ii) for each pair the
test takes $O(n\log n)$ time \cite{BrinkhoffKSS94}.
To mitigate (i), the join is evaluated in two steps.
Provided that the minimum bounding rectangles (MBRs) of the objects
are available (and possibly indexed), in the {\em filter step},
an efficient MBR-join algorithm
\cite{BrinkhoffKS93,TsitsigkosBMT19} is used to find the pairs of
objects $(r,s)\in R\times S$ such that $MBR(r)$ intersects with
$MBR(s)$.
In the {\em refinement} step, for each pair that passes the filter
step, the expensive  intersection test on the exact object geometries
is applied.
To further reduce the number of pairs needing refinement, {\em
  intermediate filters} can be added to the pipeline
\cite{BrinkhoffKSS94,ZimbraoS98}. The main idea is to
use object approximations, in addition to the MBR, that can help
to quickly determine whether a candidate pair $(r,s)$ that passes the MBR
filter is (i) a sure result, (ii) a sure non-result, or (iii) an
indecisive pair, for which we still have to apply the geometry
intersection test.
Brinkhoff et al. \cite{BrinkhoffKSS94} investigated the use of
different object approximations
(e.g., the convex hull) to be used as subsequent filters after MBR-intersection.
Zimbrao and de Souza \cite{ZimbraoS98} proposed a more effective {\em
  raster} object approximation, where each object MBR is partitioned
using a grid and the object is
approximated by the percentages of grid cell areas that the object overlaps.
This approach has several limitations. First, the raster object  representations may occupy a lot of space. Second, the approximations of two candidate objects may be based on grids of different scales; their re-scaling and
subsequent comparison can be quite expensive.
Third, the cost of comparing two rasters in order to filter a
candidate pair is linear to the number of cells in the rasters.

In a preliminary version of our work \cite{TGeorgiadis23},
we introduce {\em Raster Intervals} (RI); a raster approximation technique for polygonal objects, which does not share the drawbacks of \cite{ZimbraoS98} and reduces the end-to-end spatial join cost up to 10 times, when we use it as a pre-refinement, intermediate filter.
Our technique uses a {\em global fine grid} to approximate all objects, hence, no re-scaling issues arise. In addition, RI encodes each cell by a 3-bit sequence; whether two objects overlap in a cell can be determined by bit-wise ANDing the corresponding sequences. Finally, RI
models the set of cells that approximate an object $o$ by a sorted list of {\em raster intervals}, determined by the Hilbert curve
order of continuous cells in $o$'s representation.
For each such interval, we unify in a bitstring all 3-bit sequences of the included cells.
Object pair filtering is then implemented as a merge join between the
corresponding raster interval lists. For each pair of intersecting
intervals, the sub-bitstrings corresponding to the common cells are
ANDed to find whether there is at least one cell wherein the polygons
overlap.

Despite its effectiveness and efficiency compared to previous filters,
RI has a relatively high preprocessing cost and occupies
significant space.
In this extended version of \cite{TGeorgiadis23} we propose
\methodname{} (Approximating Polygons as Raster Interval Lists), a
significant improvement over RI.
Unlike \cite{ZimbraoS98,TGeorgiadis23} that divide the raster cells intersecting a polygon into three classes, \methodname{} uses only two cell classes, which improves storage efficiency and accelerates the intermediate filter.
Second, the main novelty of \methodname{} lies in the way it
represents objects using
{\em two lists of intervals}: 
the first (\textit{A}-list) includes all cells, regardless of their class, and the second (\textit{F}-list) includes only cells that are fully covered by the object.
The intermediate filter is then implemented as a sequence of three simple merge joins between the sorted interval lists of a given object pair.
The first join, performed between the two \textit{A}-lists, effectively identifies all true negatives. 
The last two joins, performed between one object's \textit{A}-list and the other object's \textit{F}-list, identify true positives.
Since \methodname{} does not  explicitly store or encode cell-class information and
does not perform cell-specific comparisons, it is significantly faster than previous raster approximations.
Finally, \methodname{} applies a compression technique based on delta
encoding to greatly reduce the space required to store the interval lists.
As a result, \methodname{} approximations may require even less space
than object MBRs, allowing them to be stored and processed in main memory. 
Moreover, \methodname{}'s compression scheme allows partial, on-demand
decompression of interval lists during interval join evaluation.

In addition to improving RI to \methodname{}, in this paper we show the generality
of \methodname{} in supporting spatial selection queries, spatial
within joins, and joins between polygons and linestrings. 
Furthermore, we present a space partitioning approach, which increases the
resolution of the raster grid and achieves more refined object approximations as necessary, leading to fewer inconclusive cases and,
therefore, faster query evaluation. We also investigate options for
defining and joining \methodname{} approximations of different polygons at different granularities based on their geometries.
Finally, a significant contribution of this paper is a novel, one-step ``intervalization'' algorithm
that computes the \methodname{} approximation of a polygon without having to
rasterize it in full. We show that this method
is orders of magnitude faster
compared to other rasterization approaches on CPU \cite{ZimbraoS98,Teng21}.

The rest of the paper is structured as follows: Section
\ref{sec:background} provides the necessary background.
In Section \ref{sec:ri}, we introduce our raster approximations (RI)
technique as an intermediate filter for spatial intersection joins.
Section
\ref{sec:methodology} introduces \methodname{}, our improved raster
intervals representation, detailing its features, construction, and usage. 
Section \ref{sec:customization} presents customization options for tuning \methodname{} to specific system or
dataset requirements.
In Section \ref{sec:rasterization}, we study the efficient construction  of  \methodname{} approximations.
Section \ref{sec:exp} presents our experiments that verify \methodname{}'s performance. 
Section \ref{sec:related} reviews related work, and finally, Section \ref{sec:conclusions} concludes the paper and offers suggestions for future work.

%% file: background.tex
\section{Background} \label{sec:background}
\begin{figure}[htb]
	\centering
	\includegraphics[width=\columnwidth]{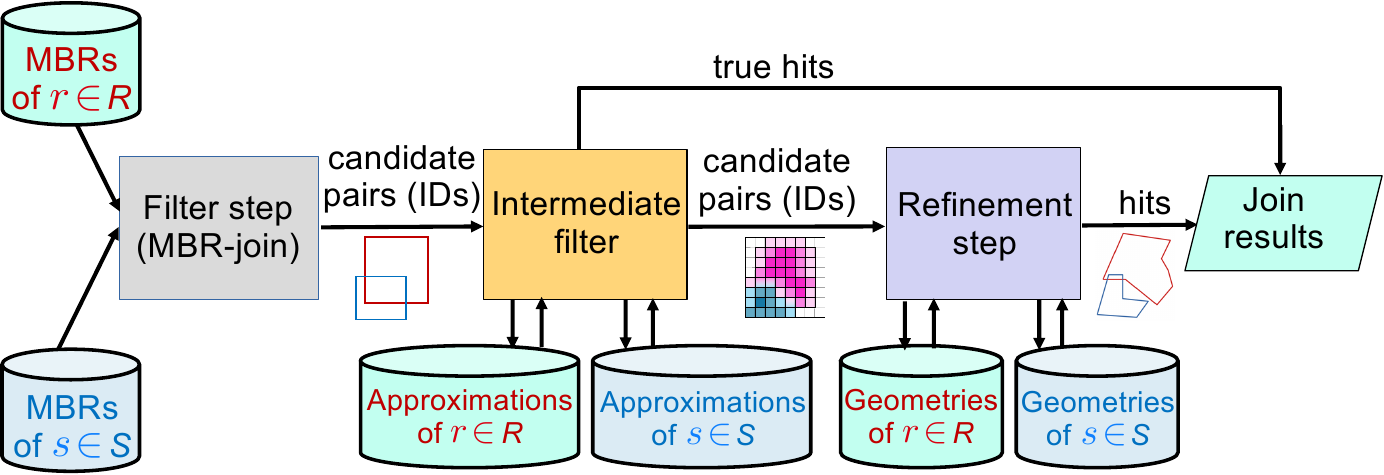}
	\caption{Spatial intersection join pipeline \cite{BrinkhoffKSS94}}
	\label{fig:pipeline}
\end{figure}

Figure \ref{fig:pipeline} illustrates the spatial intersection join pipeline. 
An MBR-join algorithm takes as input the MBR approximations of objects
to identify all pairs of objects that intersect ({\em filter step}) \cite{JacoxS07,TsitsigkosBMT19}.
Before accessing and comparing the exact object geometries for each such candidate pair, in an {\em intermediate step}, more detailed object approximations (than the MBR) are used to verify (fast) whether the pair is a sure result (true hit) or a sure non-result (false hit), or we cannot decide based on the approximations \cite{BrinkhoffKSS94,ZimbraoS98}. Finally, if the pair is still a candidate, it is passed to the {\em refinement step} where the exact geometries are accessed and an (expensive) algorithm from computational geometry \cite{ShamosH76} is run to determine whether the pair is a result. 
Most previous work
% on spatial join have
focused on
%the efficient evaluation of
the filter step \cite{BrinkhoffKS93,JacoxS07,NobariTHKBA13,TsitsigkosBMT19}. However, the refinement step dominates the overall cost, as discussed in the Introduction.
The intermediate step using additional object approximations has been proved valuable toward reducing the overal join cost \cite{BrinkhoffKSS94}.

\begin{figure}[htb]
	\centering
	\includegraphics[width=0.6\columnwidth]{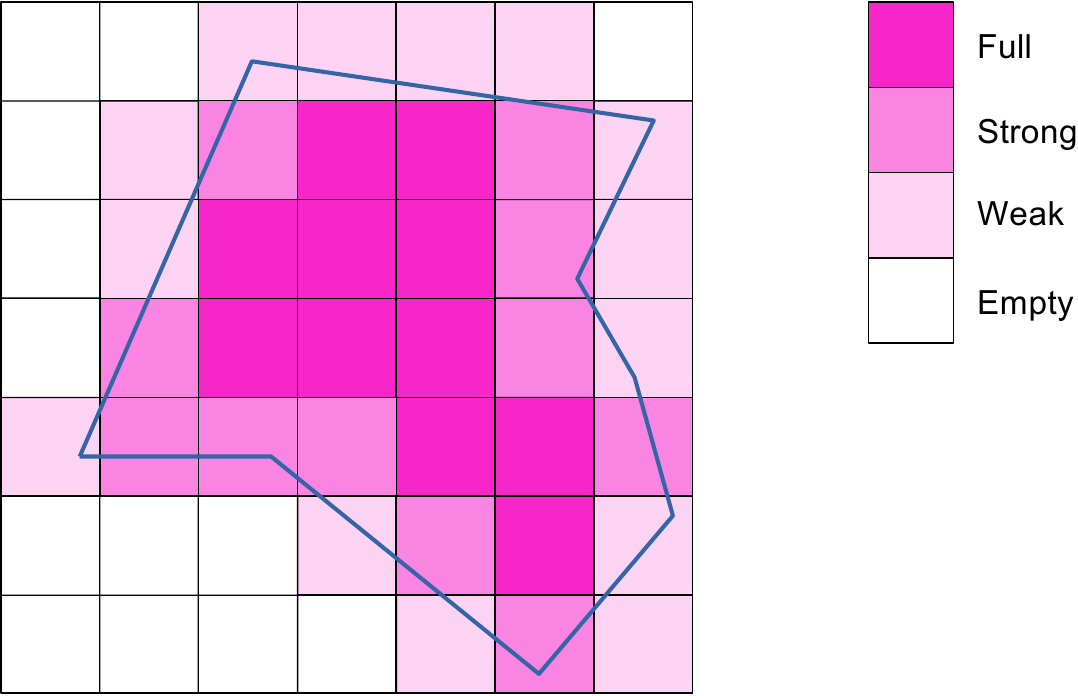}
	\caption{Four types of cells in a raster approximation \cite{ZimbraoS98}}
	\label{fig:fourclasses}
\end{figure}

%spatial join techniques consist of a filtering and a refinement step with the focus falling mostly on improving the filter \todo{example works}.
% Generally, the idea of adding another intermediate filtering step before resorting to the expensive geometric refinement has led to various methods emerging that greatly improve spatial join performance \todo{example works}. Traditionally, the filter step performs an MBR-join between the two data sets. Then, the candidate pairs pass through the intermediate filter which usually recognizes some true hits whilst the inconclusive cases are further refined geometrically. The purpose of an intermediate step is to reduce the load that needs to be refined, by approximating some results or even better by accurately recognizing them without the costly polygon-to-polygon geometric intersection. 
Zimbrao and de Souza \cite{ZimbraoS98} introduced an effective intermediate filter,
by imposing a grid over each object's MBR. The cells of the grid comprise the {\em raster approximation} of the object. Each cell belongs to one of the     
%that classifies the rasterization cells for each polygon based on the percentage of their area that is covered by the original geometry. For each polygon, the raster approximation keeps $m\times n$ cells, essentially the object's MBR, with each cell
%classified in the
following four types: \textit{full} (the object completely covers the cell), \textit{strong} (the object covers more than 50\% of the cell), \textit{weak} (the object covers at most 50\% of the cell), or \textit{empty} (the object is disjoint with the cell).
%An example of the four classes of cells in a rasterization is shown in Figure
%\ref{fig:fourclasses}.
Figure \ref{fig:fourclasses} shows an example.

To create the raster approximation (RA) of a polygon,  a grid of at most $K$ square cells is defined. The side of each cell should be $\omega 2^k$, for some $k\ge 0$, where $\omega$ is a minimum cell side (unit). In addition, the coordinates of each cell should be multiples of $\omega 2^k$.

% (Table \ref{tab:cellTypes}).
For a pair $(r,s)$ of candidate objects,
the cells in their approximations $RA(r)$ and $RA(s)$ that overlap with their common MBR are identified and the remaining ones are ignored.
If the cells of $RA(r)$ are smaller than the cells of $RA(s)$, groups of neighboring cells in $RA(r)$ are combined to infer the type of a larger cell that is perfectly aligned with a cell of $RA(s)$.
Re-scaling is expensive, results in accuracy loss and reduces the effectiveness of the filter, rendering RA useful mainly for polygons of similar size, which is rarely the case in real-world data.

After re-scaling, the common cells in the two raster approximations are examined and, for each such cell, we use the cell's types in the two approximations to
conclude whether the objects intersect in the cell,
according to Table \ref{tab:cellMatching}.
Specifically, if at least one of the two types is {\em empty}
the objects definitely do not intersect in the cell.
If at least one of the two types is {\em full} and the other is not {\em empty} or both types are {\em strong}, then the objects definitely intersect in the cell. In all other cases, we cannot conclude whether the objects intersect in the cell.
If we find at least one cell, where the objects intersect, the pair is directly reported as a spatial join result (true hit). If at all common cells, the objects do not intersect, then the pair is pruned (false hit). If we cannot conclude about the object pair, the refinement step should be applied.

%\todo{add example}

\begin{table}[htb]
	\centering
	\small
	\caption{Do two objects intersect in a cell, based on the cell's types in the two raster approximations? \cite{ZimbraoS98}}
	\label{tab:cellMatching}
	\begin{tabular}{|l|c|c|c|c|}
		\hline
		& \textbf{empty} & \textbf{weak} & \textbf{strong} & \textbf{full} \\
		\hline
			\textbf{empty} & no  &  no  & no  &  no \\ 
		\hline
		\textbf{weak} & no & {\em inconclusive} & {\em inconclusive} & yes \\ 
		\hline
		\textbf{strong} & no &  {\em inconclusive} & yes & yes\\
		\hline
		\textbf{full} & no & yes & yes & yes \\ 
		\hline
		
	\end{tabular}
%	\vspace{0.2cm}
%	\vspace{-0.4cm}
      \end{table}

%% file: ri.tex
\section{Raster Intervals}\label{sec:ri}
We propose a new framework for the intermediate step of spatial joins, which builds upon, but is significantly more effective than the raster approximation technique of previous work \cite{ZimbraoS98}. Our approach has three important differences: (i) we use the same global \rev{(and fine-grained)} grid to rasterize all objects; (ii) we use bitstring representations for the cell types of object approximations; and (iii) we represent the set of all non-empty cells of each object as a sorted list of intervals paired with binary codes. In this section, we present in detail the steps that we follow in order to generate the raster intervals approximation for each object.

\subsection{Object rasterization and raster encoding}\label{sec:RIrasterization}
We superimpose
over the entire data space (e.g., the map)
a $2^N\times 2^N$ grid.
For each data object $o$,
we identify set of the cells $C_o$ that the object intersects
and use this set to approximate $o$.
Each cell in $C_o$ may belong to three types: {\em full}, {\em strong}, or {\em weak}; as opposed to \cite{ZimbraoS98}, we do not include empty cells in $C_o$.
In order to compute $C_o$ for each object, and the type of each cell, we apply the algorithm of \cite{ZimbraoS98}. In a nutshell, the algorithm first identifies the grid columns (stripes) which overlap with $o$. It clips the object in each stripe, and then runs a plane-sweep algorithm along the stripe to identify the cells and the type of each cell.

Furthermore, we {\em encode} the three types of cells that we are using,
as shown in Table \ref{tab:codeTable}.
Note that we use a different encoding for the cell types depending on
whether the object comes from join input $R$ or $S$.
This encoding has two important properties.
First,
if for two objects $r\in R$ and $s\in S$ and for a cell $c$, the bitwise
AND of the codes of $r$ and $s$ in cell $c$ is non-zero, then
we are sure that $r$ and $s$ intersect in cell $c$.
Indeed, this corresponds to the case where at least one type
is {\em full} or both are {\em strong}. If the logical AND is 0, we cannot be sure
whether $r$ intersects $s$ in $c$.

The second property of the encoding is that it allows us to swap
the roles of $R$ and $S$ in the join, if necessary.
Specifically, the code for a cell $c$ of an object in one join input (e.g., $R$) can be converted
to the code for $c$ if the object belonged to the other join input (e.g., $S$) by XORing the code with the mask $m=110$.  For example, 011, the $R$-encoding of {\em full} cells, after bitwise XORing with $m$, becomes 101, i.e., the $S$-encoding of {\em full} cells. 
This is important for the case where the rasterization of a dataset has been {\em precomputed} before the join, according to the $R$-encoding and we want to use the dataset as the right join input $S$. \rev{XORing can be done on-the-fly when we apply our filter, as we explain in Section \ref{sec:intfilter}, with insignificant cost.}

\begin{table}[htb]
	\centering
	\small
	\caption{3-bit type codes for each input dataset}
	\label{tab:codeTable}
	\begin{tabular}{|c | c | c|} 
		\hline
		& \textbf{input $R$} & \textbf{input $S$} \\
		\hline
		\hline
		\textbf{full}   &   011 & 101 \\
		\hline
		\textbf{strong}   &   101 & 011  \\
		\hline
		\textbf{weak}   &      100 & 010   \\
		\hline
	\end{tabular}
\end{table}

\subsection{Intervalization}\label{sec:intervalizaton}
\begin{figure}[htb]
	\centering
	\includegraphics[width=0.7\columnwidth]{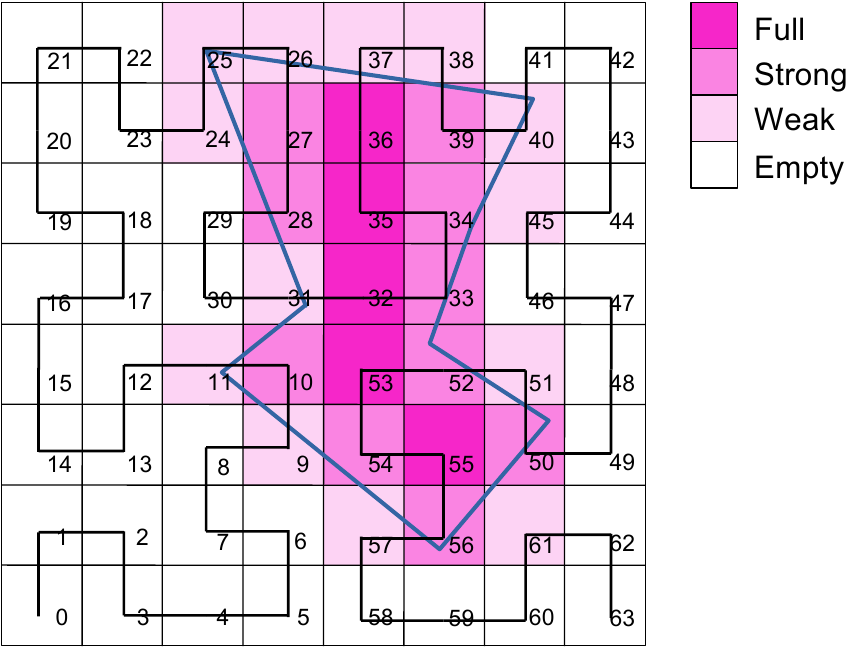}
        \includegraphics[width=0.25\columnwidth]{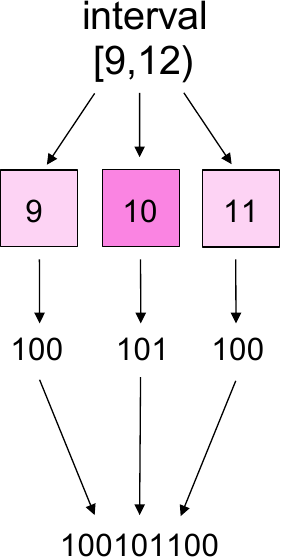}
	\caption{The Hilbert curve cell enumeration and interval generation for a polygon in a $8\times8$ space.}	
	\label{fig:intervalGeneration}
\end{figure}

We use the Hilbert curve \cite{Hilbert1891} to order the cells in the $2^N \times 2^N$ grid. Hilbert curve is a well-known space filling curve that preserves spatial proximity. Hence, each cell is mapped to a value in $[0,2^{2N}-1]$. By this, the set of cells $C_o$ that intersect an object $o$ can be represented as a list of intervals $L_o$ formed by consecutive cells in $C_o$ according to the Hilbert order. Figure \ref{fig:intervalGeneration} exemplifies the {\em intervalization} for a polygonal object $o$ in a $2^3\times 2^3$ space. The cells are marked according to their Hilbert order and shaded based on their type. There are in total 36 cells in $C_o$, which are represented by 7 intervals.
To intervalize $C_o$, we sort the cells there in Hilbert order and scan the sorted array, merging cells of consecutive cells into the current interval. The cost for this is $O(|C_o|\log|C_o|)$.

For each interval in $L_o$, during the interval construction, we {\em concatenate} the bitwise representations of the cells in their Hilbert order, to form a {\em single} code for the entire interval. This allows us to 
replace the set $C_o$ of cells that intersect an object $o$ by $L_o$.
For example, assume that the polygon of 
Figure \ref{fig:intervalGeneration} belongs to the left join input $R$.
We replace cells 9, 10 and 11 in $C_o$ with codes 100, 101 and 100, respectively, by  
interval $[9,12)$ with binary code 100101100, as shown in the figure.
This helps us to greatly reduce the space requirements for the rasterized objects.
In addition, as we will show next, we save many computations while verifying a pair of objects, because we can apply the bitwise AND for multiple cells simultaneously.
The resulting {\em raster intervals} (RI) approximation of each object is a sequence of $\langle st, end, code\rangle$ triples (ordered by $st$), where $[st,end]$ is an interval in the Hilbert curve space and $code$ is a bitstring that encodes the cell types in the interval.

\stitle{Practical considerations}
A larger value for $N$ results in a finer-grained grid and thus more accurate approximations. Moreover, a polygon rasterized with higher granularity has an increased probability to have 
completely covered cells (i.e., type {\em full}), which increases the chances of the intermediate spatial join filter to identify a true hit. At the same time, a large $N$ requires more space for storing the endpoints of the intervals in $L_o$. We choose $N=16$, which results in a grid with a fine granularity; in addition, the Hilbert order of cells (i.e., the interval endpoints) can be stored as 32-bit unsigned integers.
As each cell in an interval contributes three bits to the interval's concatenated binary code, for a $[st,end)$ interval, we need $\lceil (end-st)*3/8\rceil$ bytes to encode its cells.
We may opt to compress binary codes consisting of many bytes and the RI approximation of an object, overall.

\subsection{Intermediate filter}\label{sec:intfilter}

For a join candidate pair $(r,s)$, $r\in R, s\in S$ which is produced by the MBR-join algorithm, our objective is to use the raster intervals approximations $RI(r)$ and $RI(s)$ of $r$ and $s$ to verify fast  whether $r$ and $s$ definitely intersect, (ii) $r$ and $s$ definitely do not intersect, or (iii) we cannot conclude about the intersection of $r$ and $s$, based on their RIs.
This is done via our RI-join procedure (Algorithm \ref{alg:riajoin}).

\begin{algorithm}
	\begin{algorithmic}[1]
		\small
		\Require $RI(r)$ as $X$, $RI(s)$ as $Y$
		\State $ovl\gets False$; \Comment{no overlapping interval pair found yet}
		\State $i\gets 0$; $j\gets 0$
		\While{$i< |X|$ and $j < |Y|$} 
		  \If{$X_i$ overlaps with $Y_j$}
    		\If{\Call{AlignedAND}{$X_i.code,Y_j.code$}}
	        	\State \Return {\em true hit}\Comment{bitwise AND is non-zero}
		    \EndIf
		  \State $ovl\gets True$; \Comment{found an overlapping interval pair}
		  \EndIf   
		  \State {\bf if} $X_i.end \le Y_j.end$ {\bf then} $i\gets i+1$ {\bf else} $j\gets j+1$  
		\EndWhile
		\If{$ovl$} \Comment{at least one overlapping interval pair}
		\State \Return {\em indecisive}
		\Else  
		\State \Return {\em false hit} \Comment{no common cells in $X$ and $Y$}
		\EndIf

	\end{algorithmic}
	\caption{RI-join algorithm}
	\label{alg:riajoin}
\end{algorithm}

\begin{figure}[htb]
	\centering
	\includegraphics[width=\columnwidth]{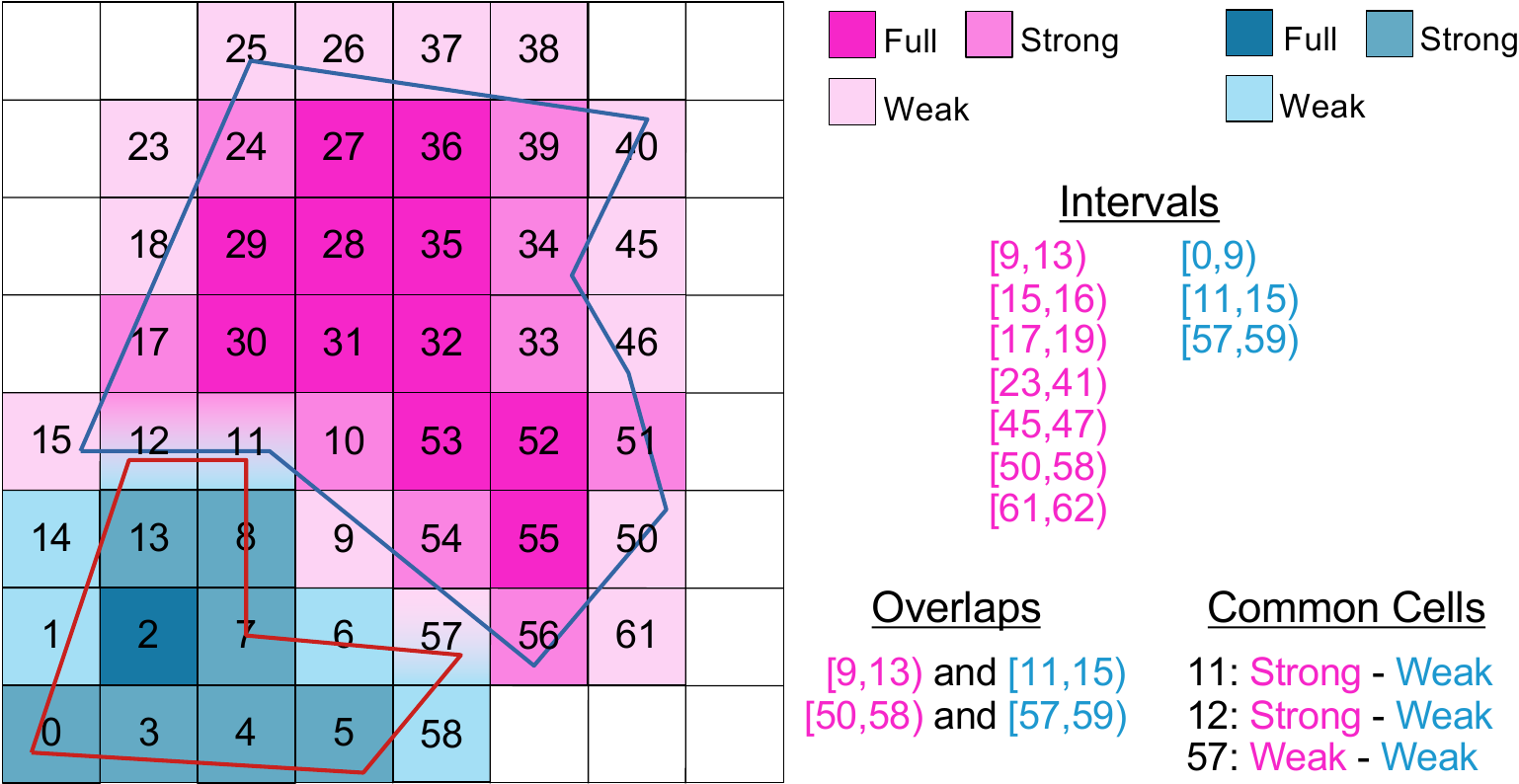}
	\caption{Two rasterized polygons, the overlaps between their raster intervals, and their common cells}
	\label{fig:intervaljoin}
\end{figure}

RI-join merge-joins the sorted interval lists $RI(r)$ and $RI(s)$, denoted by $X$ and $Y$ in the pseudocode, respectively, and identifies pairs $(X_i,Y_j)$ of intervals that overlap; i.e., $X_i$ and $Y_j$ include at least one common cell.
For each such pair, there is a possibility to find out that $(r,s)$ is a true hit (i.e., a spatial join result) and avoid sending the pair to the refinement step.
Specifically, if in at least one of the common cells of $X_i$ and $Y_j$ the logical AND of the cell codes is non-zero, we have a sure true hit and we do not need to continue the RI-join. Having the codes of the cells in $X_i$ and $Y_j$ concatenated in two single bitstrings  $X_i.code$ and $Y_j.code$ allows us to perform this check (abstracted by Function {\sc AlignedAND}) efficiently. We first select from each bitstring the  fragment that includes the codes of all cells in $[\max\{X_i.st,Y_j.st\}, \min\{X_i.end,Y_j.end\}]$, i.e., the intersection interval of  $X_i$ and $Y_j$. Then, we bitwise AND the fragments.
If the fragments have been encoded by the same encoding (i.e., both have $R$ or $S$ encoding as shown in Table \ref{tab:codeTable}),
ANDing is preceded by XORing one of the two codes.
If there is at least one  pair $(X_i,Y_j)$ of overlapping intervals
(variable $ovl$ of Algorithm \ref{alg:riajoin} is True at the end of the while-loop), but the object pair is not found to be a true hit, then the object pair is {\em indecisive}, meaning that we will have to apply the refinement step for it.
On the other hand, if there are no overlapping intervals in the two RIs ($ovl$ remains False), there are no common cells in the raster representations of the objects, and we can conclude that the two objects definitely do not intersect (false hit). 
As an example, Figure \ref{fig:intervaljoin} shows two rasterized polygons and the pairs $(X_i,Y_j)$ of intervals from the two raster intervals that overlap.

In general, the codes (bitstings) of two intersecting intervals may occupy multiple bytes and the common subinterval may be of arbitrary length. Before bit-shifting, Function {\sc AlignedAND} truncates all un\-ma\-tched bytes from the two bitstrings. In addition, bit-shifting is done at the bytes of one interval only (the one that starts earlier), making sure to carry over the required bits from the next byte to avoid any loss of information. This continuous shifting and matching (binary AND between aligned bitstrings) is performed byte-by-byte, hence, once two ANDed bytes give a non-zero, we immediately report the true hit. XORing, (if both join inputs have the same encoding), is done on-demand on the shifted byte, after any potential bit carryover. A byte-wide XOR mask $m_{byte}$ is used, created by concatenating our mask $m=110$ a few times to fill a byte; $m_{byte}$ is shifted, if necessary.
The whole process can easily be parallelized (shifting and bitwise operations are independent for each byte).

For each pair of intervals, the last bytes to be matched is a special case and has to be treated cautiously, since the remaining bits that need checking may be less than 8  and the rest of the bits in that byte should not be included in the bitwise operations.
In other words, the XOR and AND operations applied on the last bytes should consider bits only in the positions relevant to the compared intervals, otherwise we may mistake a false positive as a true hit. Hence, we apply one last bit mask with 1s at the positions of the bits that need to partake in the operation, setting the rest to zero.

\begin{figure}[htb]
	\centering
	\includegraphics[width=\columnwidth]{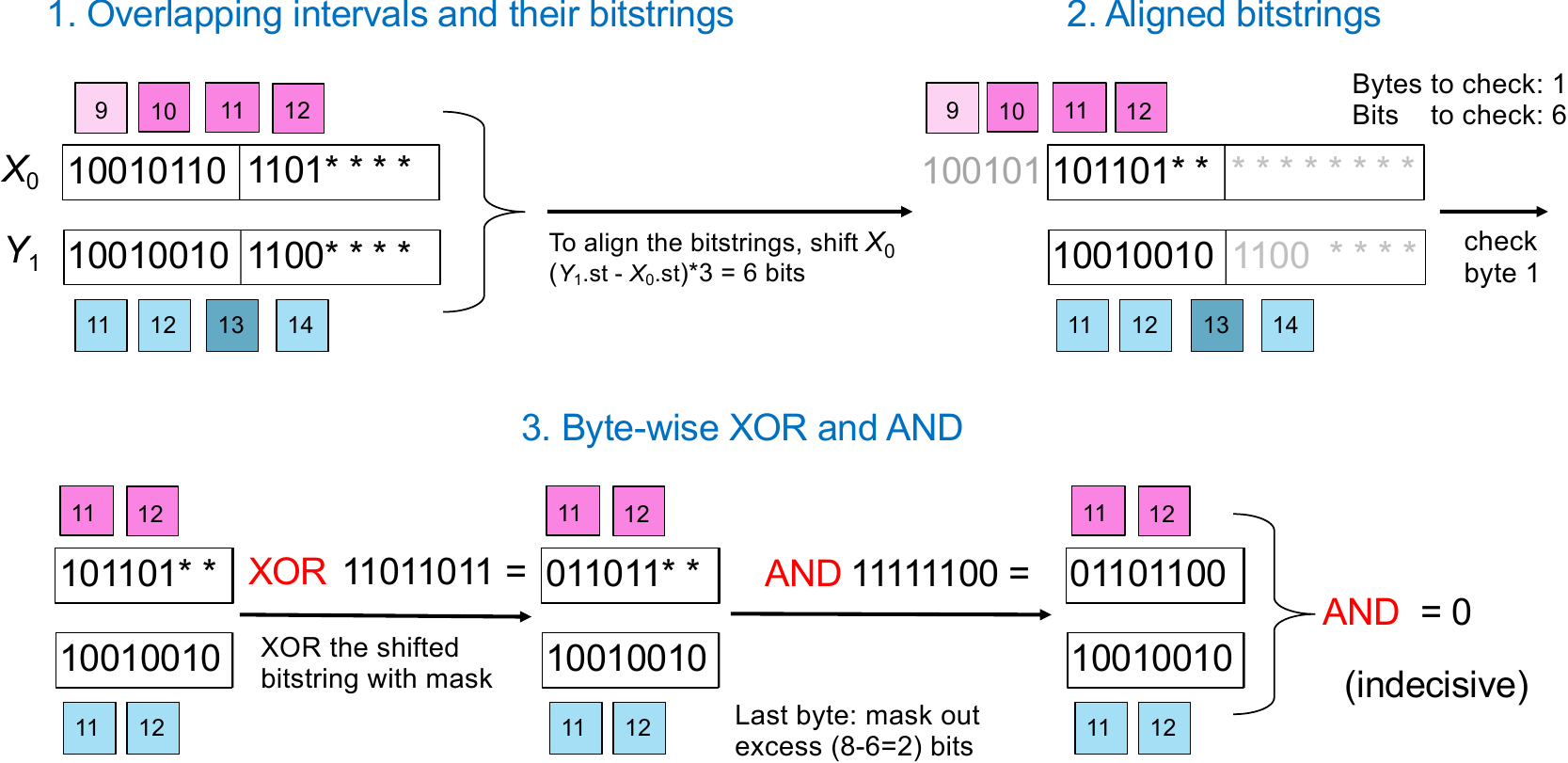}
	\caption{Intervals $[9,13)$ and $[11,15)$ of our two example polygons overlap but are not aligned. Byte truncation and bit shifting (if necessary) align their bitstrings before performing the bitwise operation(s)}
	\label{fig:bitShifting2}
      \end{figure}

Figure \ref{fig:bitShifting2} shows how the codes for first pair  $(X_0,Y_1)$ of intersecting intervals from the example of  Figure \ref{fig:intervaljoin} are matched,
where $X_0=\langle [9,13),100101101101\rangle$ and
$Y_1=\langle [11,15),100100101100\rangle$ (i.e., assume that both datasets are $R$-coded).
Each code occupies 2 bytes.
Since the interval of $Y_1$ starts $2$ cells after the interval of $X_0$, the code of $X_0$ is shifted by $2\times 3=6$ bits in the first step. This aligns the common cells (11 and 12) in the two codes.
The common fragment (6 bits) occupies 1 byte, so there will be one byte-by-byte match.
As both intervals are $R$-coded, we first XOR the $X_0$-byte with the (shifted) byte-wise XOR mask $m_{byte}$. Before ANDing the two bytes, we AND the shifted byte with a mask that clears the bits that are outside the common fragment of the intervals, as we are at the last byte. Finally, the bytes are ANDed with a 0 result, so the intersection of the two objects remains indecisive with respect to  $(X_0,Y_1)$.
As a result, Algorithm \ref{alg:riajoin} continues to find next pair of overlapping intervals $(X_5,Y_2)$ and performs the corresponding code matching.

\stitle{Analysis.}
RI-join requires a single scan of interval lists $X$ and $Y$, since no two intervals in the same list (i.e., in the same polygon) overlap.
Assuming that bitstrings are relatively short so that their matching (a call to Function {\sc AlignedAND}) takes constant time,
the time complexity of Algorithm \ref{alg:riajoin}  is $O(|X|+|Y|)$ since the number of overlapping interval pairs is at most $|X|+|Y|$.

\subsection{``Within'' spatial joins}\label{sec:within}
Although we focus on polygon-polygon intersection joins,
RI can also be used as an intermediate filter for {\em within joins}.
The objective of a spatial within join is to find pairs $(r,s)$ of objects, $r\in R$, $s\in S$, such that $r$ is {\em within} $s$, i.e. the space occupied by $r$ is a subset of the space occupied by $s$. 
For each pair $(r,s)$ of polygons that passes the filter step of the within join (i.e., the MBR of $r$ is within the MBR of $s$), we can apply Algorithm \ref{alg:riajoin}
with the following changes in order to identify whether $(r,s)$ is a true negative (false hit), a true positive (i.e., true hit), or an indecisive pair w.r.t. the within predicate:
As soon as we find an interval $X_i \in RI(r)$ which is not a subset of any interval  $Y_j \in RI(r)$, we can terminate with the assertion that $r$ is not within $s$, since there is at least one non-empty cell of $r$ which is empty in $s$.
In addition,
for an identified pair of $(X_i, Y_j)$, such that $X_i\subseteq Y_j$, if there is a cell in $X_i$ that is (i) {\em full} in $X_i$ but not full in $Y_j$ or (ii) {\em strong} in $X_i$ and {\em weak} in $Y_J$, then $(r,s)$ should a true negative and the algorithm terminates.
For $(x,y)$ to be characterized as a true hit without refinement, for all identified $(X_i,Y_j)$ such that $X_i\subseteq Y_j$, all cells in the subinterval $X_i$ where $X_i$ and $Y_j$ overlap should be {\em full} in $Y_j$; if at least one such cell is not full, then we cannot guarantee a true hit and
the pair $(x,y)$ must be passed to the refinement step, unless it is found to be a true negative.

%% file: methodology.tex
\section{APRIL}\label{sec:methodology}

We now propose
\textit{\methodname{}} (Approximating Polygons as Raster Interval Lists), a significant enhancement of RI, which can be used as an intermediate filtering method for spatial query processing and is more efficient and less space consuming compared to
RI.

\subsection{\textit{A}- and \textit{F}-Interval Lists}\label{subsec:intervallists}
\methodname{} is a succinct polygon approximation for intermediate filtering, which categorizes raster cells into {\em Full}, {\em Partial}, and {\em Empty}, based on their coverage percentage with the object's geometry ($100\%$, less than $100\%$, and $0\%$, respectively).
In other words,
\methodname{} {\em unifies} the {\em Strong} and {\em Weak}  cell classes used by RI and \cite{ZimbraoS98}  to a single {\em Partial} class.
Under this, \methodname{}  approximates a polygon with two sorted interval lists:
the \textit{A}-list and the \textit{F}-list.
The \textit{A}-list contains intervals that concisely capture all cells that overlap with the polygon, regardless of their type (Full or Partial),
whereas the \textit{F}-list includes only Full cells.
An interval list having $n$ intervals is stored as a simple sorted integer sequence in which the $i$-th interval's $start,end$ are located at positions $2i$ and $2i + 1$ respectively, for $i \in [0,n)$. 
\begin{figure}[htb]
	\centering
	\includegraphics[width=\columnwidth]{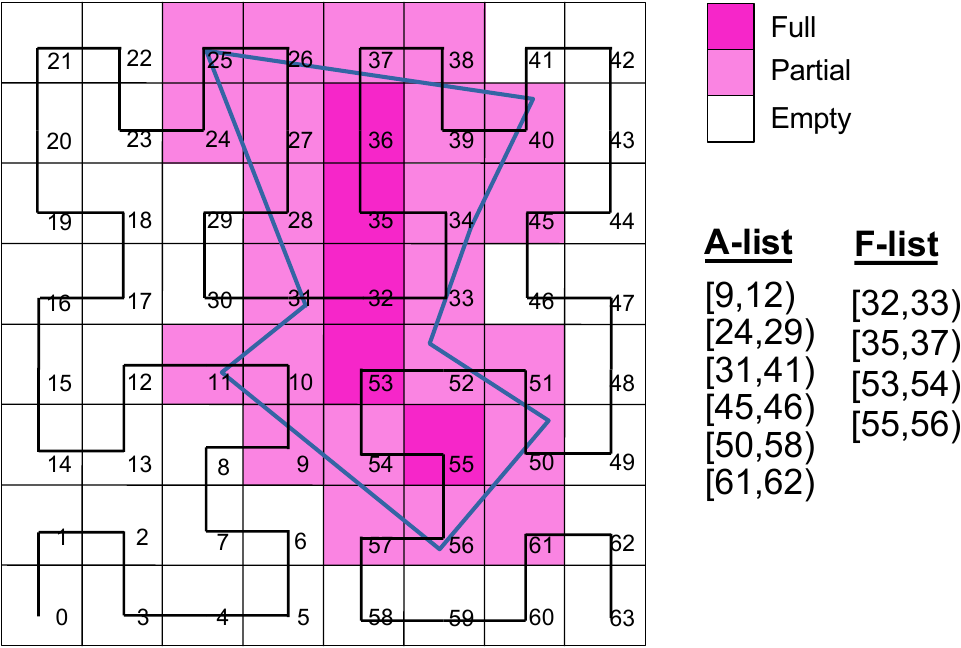}
	\caption{The interval generation for a polygon in a $8\times8$ space, without bit-coding and using interval lists.}
	\label{fig:intervalLists}
\end{figure}

The \textit{A}-list and \textit{F}-list for the example polygon of Figure \ref{fig:intervalGeneration} are shown in Figure \ref{fig:intervalLists}.
Strong and Weak cell types become Partial, which results in a simpler representation than \oldmethod{}. 
Note that the set of intervals in each of the \textit{A}- and \textit{F}- lists are disjoint. 
The new relationship identification table for a cell shared by two polygons, is shown in Table \ref{tab:newCellMatching}.
Removing the Strong cell type renders the approximation unable to detect true hits for cells of the Strong-Strong case, as common cells that are both Partial cannot decide definite intersection between the two polygons.%
\footnote{As we have found experimentally  (Sec.  \ref{subsec:spatialjoinsexperiments}),
this has minimal effect on the amount of true hits and true negatives that the intermediate filter manages to detect.
This is due to the fact that the only cases of true hits missed are pairs of polygons that intersect with each other \textit{exclusively} in cells typed Strong for both polygons and nowhere else.}

\begin{table}[htb]
	\centering
	\caption{\methodname{}: Do two objects intersect in a common cell?}
	\label{tab:newCellMatching}
	\vspace{-0.2cm}
	\small	
	\begin{tabular}{|l|c|c|}
		\hline
		& \textbf{Partial} & \textbf{Full} \\
		\hline
		\textbf{Partial} & {\em Inconclusive} & yes \\
		\hline
		\textbf{Full} & yes & yes \\ 
		\hline
		
	\end{tabular}
\end{table}

\stitle{Construction}
To construct an \methodname{} approximation
we need to first identify the cells intersected by the polygon's area in the grid, while also labeling each one of them as Partial or Full.
Then, \textit{Intervalization}
derives the \textit{F-}list, by sorting the set of Full cells by ID (i.e., Hilbert order) and merging consecutive cell IDs into intervals.
To derive the \textit{A-}list, we repeat this for the union of Full and Partial cells.
In Section \ref{subsec:onesteprasterization}, we propose an efficient algorithm that derives the \textit{F-} and \textit{A-}list of a polygon without having to label each individual cell that intersects it.

\subsection{\methodname{} Intermediate Spatial Join Filter}\label{subsec:spatialjoins}

\methodname{} is used as an intermediate filter (Figure \ref{fig:pipeline}) that is situated between the MBR filter and the refinement phase.
Given a pair $(r,s)$ of objects coming as a result of an MBR-join algorithm \cite{BrinkhoffKS93,PatelD96,TsitsigkosBMT19}, \methodname{} uses the \textit{A-}  and \textit{F-}lists of $r$ and $s$ to detect fast
whether the polygons (i) are disjoint (true negative), (ii) are guaranteed to intersect (true hit), or (iii) are inconclusive, so they have to be forwarded to the refinement stage to verify their intersection. 

Whether $r$ and $s$ are disjoint (i.e. do not intersect), can be determined by checking whether their \textit{A-}lists have any pair overlapping of intervals or not.
If they have no overlapping intervals, then  $r$ and $s$  do not have any common cell in the grid and thus they cannot intersect. \revj{We check this condition by
performing a merge join over the} \textit{A}-lists and stopping as soon as we detect two overlapping
intervals.

Pairs of polygons that have at least one pair of overlapping intervals in their \textit{A}-lists are then checked using their \textit{F}-lists. We perform two more merge-joins: $\textit{A(r)} \bowtie \textit{F(s)}$ and $\textit{F(r)} \bowtie \textit{A(s)}$;
detecting an overlapping intervals pair in one of these two joins means that there is a Full cell in one object that is common to a Full or Partial cell of the other object. This guarantees that the two objects intersect and the pair $(r,s)$ is immediately reported as a spatial join result. If $\textit{A(r)} \bowtie \textit{F(s)}$ fails to detect $(r,s)$ as a true hit, then $\textit{F(r)} \bowtie \textit{A(s)}$ is conducted; if the latter also fails, then $(r,s)$ is an {\em inconclusive} candidate join pair, which is forwarded to the refinement step. 

\begin{figure}[htb]
	\centering
	\includegraphics[width=\columnwidth]{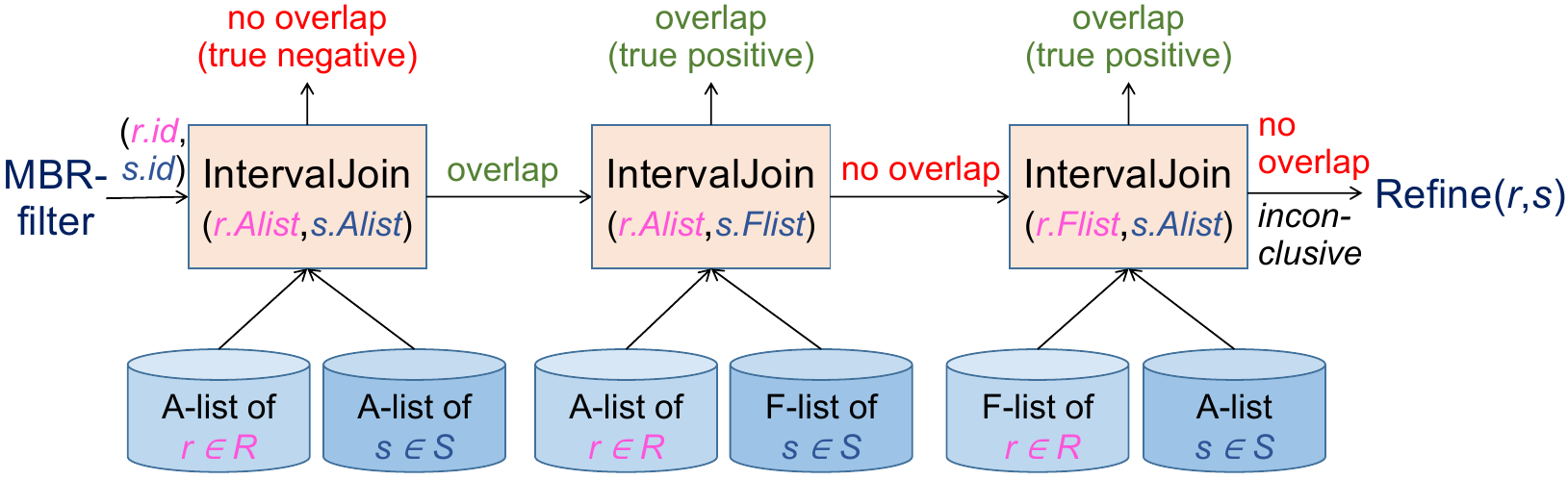}
	\vspace{-4ex}
	\caption{The three steps of the intermediate filter for a candidate pair of polygons.}	
	\label{fig:intermediateFilter}
\end{figure}

In summary, \methodname{}'s intermediate filter sequence consists of three steps: the \textit{AA}-join, \textit{AF}-join, and \textit{FA}-join, as illustrated in Figure \ref{fig:intermediateFilter} and described by
Algorithm \ref{alg:apriljoin}.
Each step is a simple merge-join between two sorted interval lists. Since each list contains disjoint intervals, each of the three interval joins takes $O(n+m)$ time, where $n$ and $m$ are the lengths of the two interval join input lists. Hence, the total cost of the \methodname{} filter (i.e., Algorithm \ref{alg:apriljoin}) is linear to the total number of intervals in the  \textit{A}-  and \textit{F}-lists of $r$ and $s$.

\begin{algorithm}
	\footnotesize
	\begin{algorithmic}[1]
		\Require{$(r,s)$ such that $MBR(r)$ intersects $MBR(s)$}
		\Function{IntervalJoin}{$X, Y$}	
		\State $i\gets 0$; $j\gets 0$
		\While{$i< |X|$ and $j < |Y|$} 
		\If{$X_i$ overlaps with $Y_j$}
		\State \Return \textit{true}\Comment{overlap exists}
		\EndIf
		
		\State {\bf if} $X_i.end \le Y_j.end$ {\bf then} $i\gets i+1$ {\bf else} $j\gets j+1$  
		\EndWhile
		
		\State \Return \textit{false}\Comment{no overlaps detected}
		\EndFunction
		\\
		
		\If{not IntervalJoin($A(r), A(s)$)}
		\State \Return \textit{false} \Comment{true negative}
		
		\EndIf
		
		\If{IntervalJoin($A(r), F(s)$)}
		\State \Return \textit{true} \Comment{true hit}
		
		\EndIf
		
		\If{IntervalJoin($F(r), A(s)$)}
		\State \Return \textit{true} \Comment{true hit}
		
		\EndIf
		
		\State \Return {REFINEMENT$(r,s)$} \Comment{forward pair to refinement}

	\end{algorithmic}
	\caption{\methodname{} join algorithm.}
	\label{alg:apriljoin}
\end{algorithm}

\stitle{Join Order Optimization}
The \textit{AA}-join, \textit{AF}-join, and \textit{FA}-join could be applied in any order in
Algorithm \ref{alg:apriljoin}.
For example, if $(r,s)$ is a true hit, it would be more beneficial to perform the \textit{AF}-join and the \textit{FA}-join before the \textit{AA}-join, as this would identify the hit earlier.
On the other hand, if $(r,s)$ is a true negative, conducting the \textit{AA}-join first avoids the futile \textit{AF}- and \textit{FA}-joins.
However, there is no way to know a priori whether $(r,s)$ is a true hit or a true negative. In addition, we experimentally found that changing the join order does not have a high impact on the intermediate filter cost and the overall cost.
For a typical candidate pair $(r,s)$ the common cells are expected to be few compared to the total number of cells covered by either $r$ or $s$, making \textit{AA}-join the most reasonable join to start with. This is confirmed by our experiments where the number of candidate pairs identified as true negatives is typically much larger compared to the number of identified true hits. 

\subsection{Generality}\label{subsec:generality}
In this section, we demonstrate the generality of \methodname{} in supporting other queries besides spatial intersection joins between polygon-sets. We first show how  we can use it as an intermediate filter in selection (range) queries. Then, we discuss its application in spatial {\em within} joins.
Finally, we discuss the potential of using \methodname{} approximations of polygons and raster approximation of linestrings to filter pairs in polygon-linestring intersection joins.

\subsubsection{Selection Queries}\label{subsec:selection}

Similarly to joins, \methodname{} can be used in an intermediate filter to reduce the cost  of selection queries. Consider a spatial database system, which manages polygons and where the user can draw a selection query as arbitrary polygon $QP$; the objective is to retrieve the data polygons that intersect with the query polygon $QP$. Assuming that we have pre-processed all data polygons and computed and stored their  \methodname{} representations, we can process polygonal selection queries as follows. We first pre-process $QP$ to  create its \methodname{} approximation. Then, we use the MBR of $QP$ to find fast the data polygons whose MBR intersects with the MBR of the query (potentially with the help of an index \cite{AGuttman84,DTsitsigkos21}).
For each such data polygon $r$,
we apply the \methodname{}  intermediate filter for the $(r,QP)$ pair to find fast whether $r$ is a true negative or a true hit. If $r$ cannot be pruned or confirmed as a query result, we eventually apply the refinement step. 

\subsubsection{Spatial Within Joins}\label{subsec:within}

\methodname{}  can also applied for spatial joins having a \textit{within} predicate, where the objective is to find the pairs $(r,s)$, where $r \in R$ and $s \in S$ and $r$ is \textit{within} $s$ (i.e., $r$ is completely covered by $s$). In this case, the intermediate filter performs only two of its three steps.
The \textit{AA}-join is applied first to detect whether $r$ and $s$ are disjoint, in which case the pair should be eliminated.
Then, we perform a variant of the \textit{AF}-join, where the objective is to find if {\em every} interval in the \textit{A}-list of $r$ is contained in one interval in the
\textit{F}-list of $s$; if this is true, then $(r,s)$ is guaranteed to be a within join result and it is reported as a true hit. In the opposite case, $(r,s)$ is forwarded to the refinement step.
We do not apply an
\textit{FA}-join, because this may only detect whether $s$ is within $r$.

\subsubsection{Linestring to Polygon Joins}\label{subsec:linestring}
\begin{figure}[htb]
	\centering
	\includegraphics[width=0.8\columnwidth]{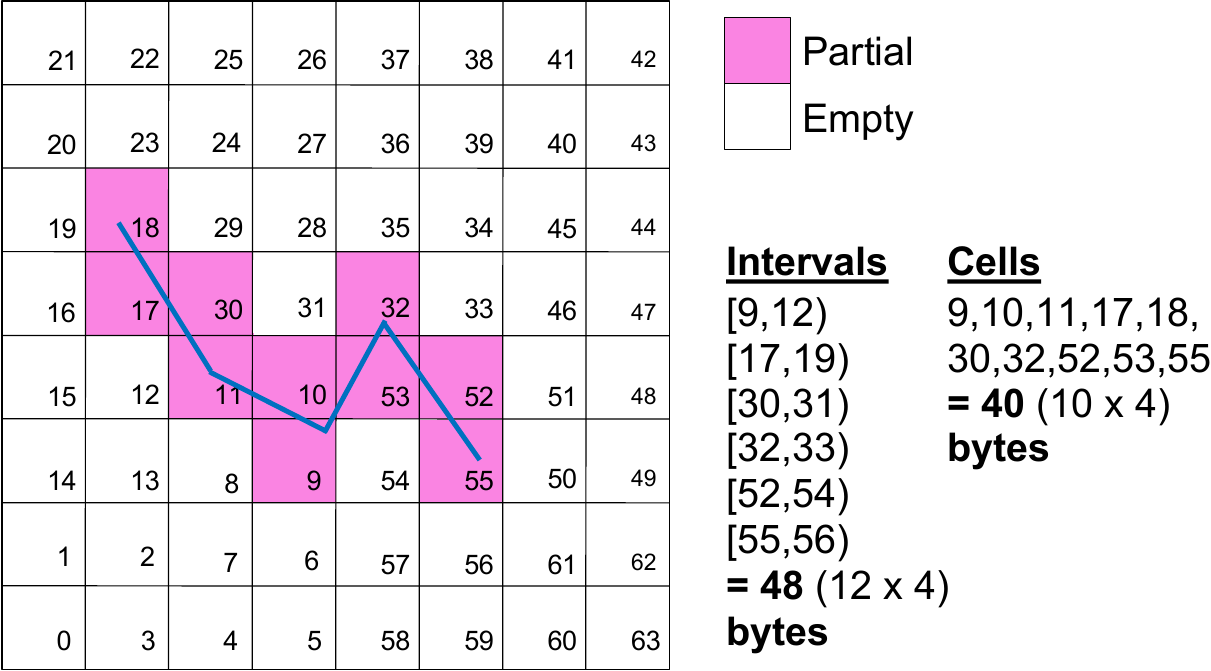}
	\vspace{-2mm}
	\caption{A linestring's \methodname{} approximation size in bytes, if stored as intervals versus cells.}	
	\label{fig:linestrings}
\end{figure}

Another interesting question is whether \methodname{} can be useful for intersection joins between other spatial data types, besides polygons.
The direct answer is no, since \methodname{} is designed for spatially-extended objects.
Still, our method can be useful for the case of joins between polygons and linestrings.
A linestring is a sequence of line segments and it is used to approximate geographic objects such as roads and rivers. 
The rasterization of a linestring results in only Partial cells, as linestrings have zero area and cannot cover a cell entirely.
In addition, as exemplified in Figure \ref{fig:linestrings}, linestrings do not really benefit from merging consecutive cells into intervals, as linestrings that follow the Hilbert order (or any other fixed space-filling curve) are rare.
Hence, it is more space-efficient to approximate a linestring as a sorted sequence of cell-IDs (which are guaranteed to be Partial). 
Having the linestring approximations, we can evaluate spatial intersection joins between a collection of polygons and a collection of linestrings,
by applying two of the three steps in the \methodname{} intermediate filter; namely,
(i) a merge-join between the \textit{A}-list of the polygon and the  cell-ID list of the linestring to find out whether the pair is a true negative and (ii) 
a merge-join between the \textit{F}-list of the polygon and the cell-ID list of the linestring to find out whether the pair is a true hit. Algorithm \ref{alg:apriljoin} can easily be adapted for polygon-linestring filtering, by changing IntervalJoin($X, Y$) to take a sequence of cell-IDs $Y$ and treat them as intervals of duration 1.

%% file: customization.tex
\section{Customization}\label{sec:customization}

We have explored a series of optimization and customization options that can potentially reduce \methodname{}'s space complexity and improve its performance in terms of filter effectiveness and speed.

\subsection{Compression}\label{subsec:compression}

Recall that the only information that
\methodname{} stores for each polygon is two interval lists:
the \textit{A}-list  and the \textit{F}-list.
The interval lists are essentially sorted integer arrays, so we can exploit
delta encoding and more specialized lossless compression schemes to reduce their space requirements.
Since any of the \textit{AA}- \textit{AF}- and \textit{FA}-join that we may apply on the lists may terminate early (as soon as an interval overlap is detected), we should go for a compression scheme that does not require the decompression a list entirely before starting processing it. In other words, we should be able to perform joins \textit{while} decompressing the lists. This way, we may avoid uncompressing the lists at their entirety and still be able to perform the joins. In view of this, we use delta encoding, where we store the first value of the list precisely and from thereon store the differences (gaps) between consecutive numbers.

There are dozens of different compression schemes for gaps between ordered integers, each with their pros and cons. We chose the Variable Byte (VByte) method \cite{CuttingP90,ThielH72}, a popular technique that even though it rarely achieves optimal compression, it is adequately efficient and really fast \cite{Lemire12}. 
We use the libvbyte \cite{Libvbyte} library that has an option for sorted integer list compression, which matches our case and boosts performance by utilizing delta encoding. Compression hardly affects \methodname{}'s construction time, which is dominated by the rasterization/intervalization cost.

At the same time, we adapt our interval join algorithm to apply decompression and join at the same time, i.e., each time it needs to get the next integer from the list it decompresses its value and adds it to the previous value in the list.

\subsection{Partitioning}\label{subsec:partitioning}

The accuracy of \methodname{} as a filter is intertwined with the grid granularity we choose. A more fine-grained grid results in more Full cells, increasing the chance of detecting true hits; similarly, empty cells increase,
enhancing true negative detection.
However, simply raising the order $N$ is not enough to improve performance.
Increasing $N$ beyond 16
means that a single unsigned integer is not enough to store a Hilbert curve's identifier, which range from $[0,2^{2N} - 1]$.
For $N=17$ or higher, we would need 8 bytes (i.e., an unsigned long) to store each interval endpoint, exploding the space requirements and the access/processing cost.

\begin{figure}[htb]
	\centering
	\includegraphics[width=0.6\columnwidth]{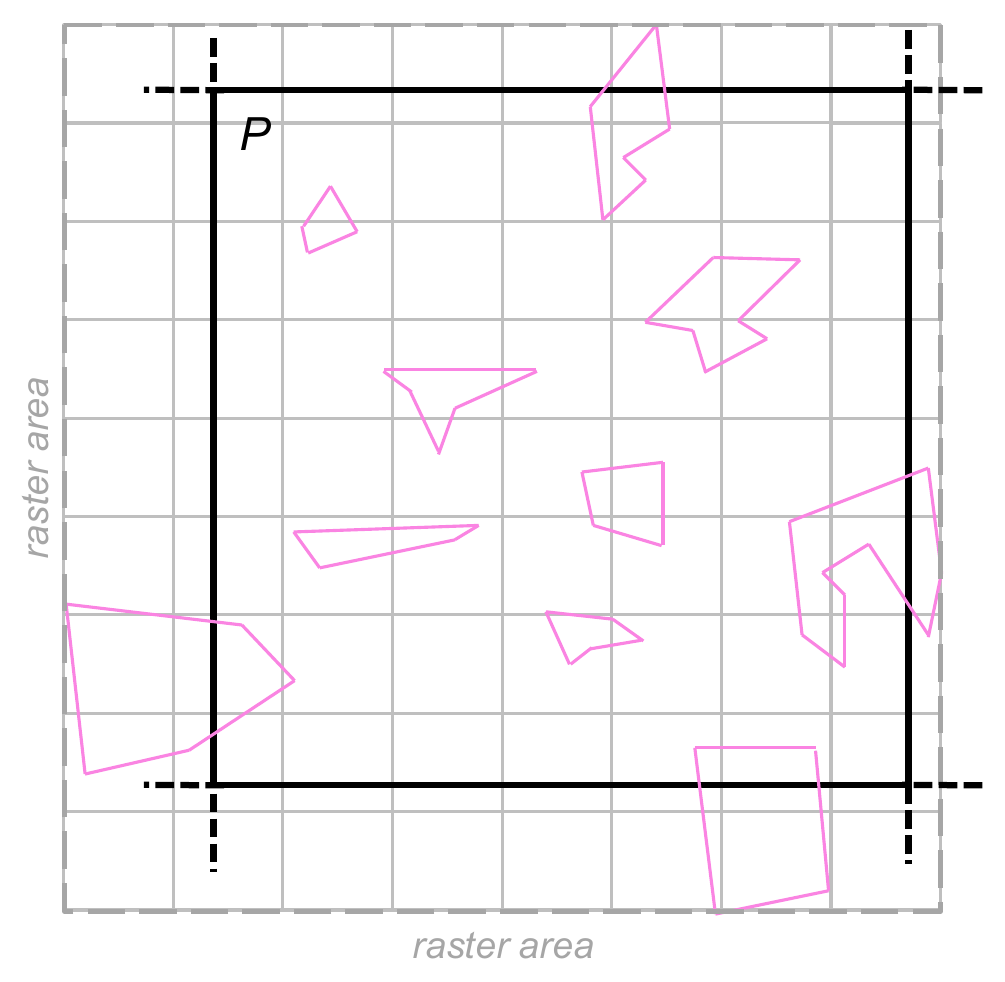}
	\vspace{-4ex}
	\caption{Example of a partition $P$, a group of polygons in it and $P$'s raster area with granularity order $N=8$.}	
	\label{fig:partitionareas}
\end{figure}

In view of this, we introduce a partitioning mechanism for \methodname{}, that divides the data space into {\em disjoint} partitions and defines
a dedicated rasterization grid and Hilbert curve of order $N=16$ to each partition.
This increases the global granularity of the approximation, without using long integers, while giving us the opportunity to define smaller partitions for denser areas of the map for which a finer granularity is more beneficial.
Partitioning is done considering all datasets (i.e., layers) of the map. That is, the same space partitioning is used for all datasets that are joined together.
The contents of each partition are all objects that intersect it; hence, the \textit{raster} area of the partition is defined by the MBR of these objects and may be larger than the partition, as shown in the example of Figure \ref{fig:partitionareas}.
\methodname{} approximations are defined based on the raster area of the partition.
The spatial join is then decomposed to multiple joins, one for each spatial partition. Duplicate join results are avoided at the filter step of the join (MBR-join) as shown in \cite{Dittrich00,TsitsigkosBMT19}.
\revj{This partitioning approach is particularly beneficial for big data management systems, such as Apache Sedona, where spatial queries are evaluated in a distributed manner after space and data partitioning.
Partitioning allows (i) more accurate approximations through fine-grained partitions and (ii) parallel evaluation per partition that further reduces the end-to-end join time.}

\subsection{Different Granularity}\label{subsec:differentGranularity}
 
If we use the same (fine) grid to rasterize all polygons, the \methodname{} approximations of large polygons may contain too many intervals, slowing down the intermediate filter.
We can create approximations using a different order $N$ of the Hilbert curve for different datasets, based on the average sizes of their contents. There is a trade-off between memory and performance, since an order lower than $16$ means fewer intervals and thus lower memory requirements and complexity, but also means reduced \methodname{} accuracy.

When joining two \methodname{} approximations of different order, we need to adjust one of the two interval lists so that it can be joined with the other.
For this, we
scale down the list with the highest order.
Specifically, before comparing two intervals $a=[a_{start}, a_{end})$ and $b=[b_{start}, b_{end})$ at orders $N$ and $L$ respectively, where $N>L$, the highest order interval $a$ should be right shifted by $n = |N-L|\times 2$ bits, to form a transformed interval $a'$, as follows:

\begin{equation}
a' = [a_{start} >> n, (a_{end}-1) >> n]
\end{equation}

Right shifting creates intervals in a more coarse-grained grid and thus, they may represent larger areas than the original. Therefore, this formula works only for \textit{A}-intervals, since there is no guarantee that a Full interval at order $N$ will also be Full at order $L$.
For this reason, in Algorithm \ref{alg:apriljoin}, we perform
only one of 
the {\em AF}- and {\em FA}- joins, using the {\em F}-list of the coarse approximation (which is not scaled down).
This has a negative effect on the filter's effectiveness, as a trade-off for the coarser (and smaller) \methodname{} approximations that we may use for large polygons. 

%% file: rasterization.tex
\section{\methodname{} Approximation Construction}\label{sec:rasterization}
In this section, we present two methods for the construction of  a polygon's \methodname{} approximation.
In Section \ref{subsec:rasterization} we present a {\em rasterization} approach that
efficiently finds the cells that intersect an input polygon and their types, based on previous research on polygon rasterization, and then sorts them to construct the {\em A}- and {\em F}-interval lists. In Section \ref{subsec:onesteprasterization}, we propose a more efficient approach tailored for \methodname{}, which avoids classifying all cells, but directly identifies the intervals and constructs the {\em A}- and {\em F}-interval lists.

\subsection{Efficient Graphics-Inspired Rasterization}\label{subsec:rasterization}

RI and the previous raster-based filter of \cite{ZimbraoS98} require the classification of each cell to Full, Strong, Weak, or Empty, 
based on the percentage of the cell covered by the original polygonal geometry. 
For this, they apply an algorithm that involves numerous polygon clippings and polygonal area computations, at a high cost.
On the other hand, to define an  \methodname{} approximation,
we only need to identify the cells which are partially or fully covered by the input polygon's area.
Inspired by rasterization techniques in the graphics community
\cite{AmanatidesW87,Smith79}, we propose a polygon rasterization technique which involves two stages. Firstly, we compute the Partial cells, which essentially form the boundary of the polygon in the grid. Next, we compute the Full cells using the previously-computed boundary cells.

\begin{figure}[htb]
	\centering
	\includegraphics[width=\columnwidth]{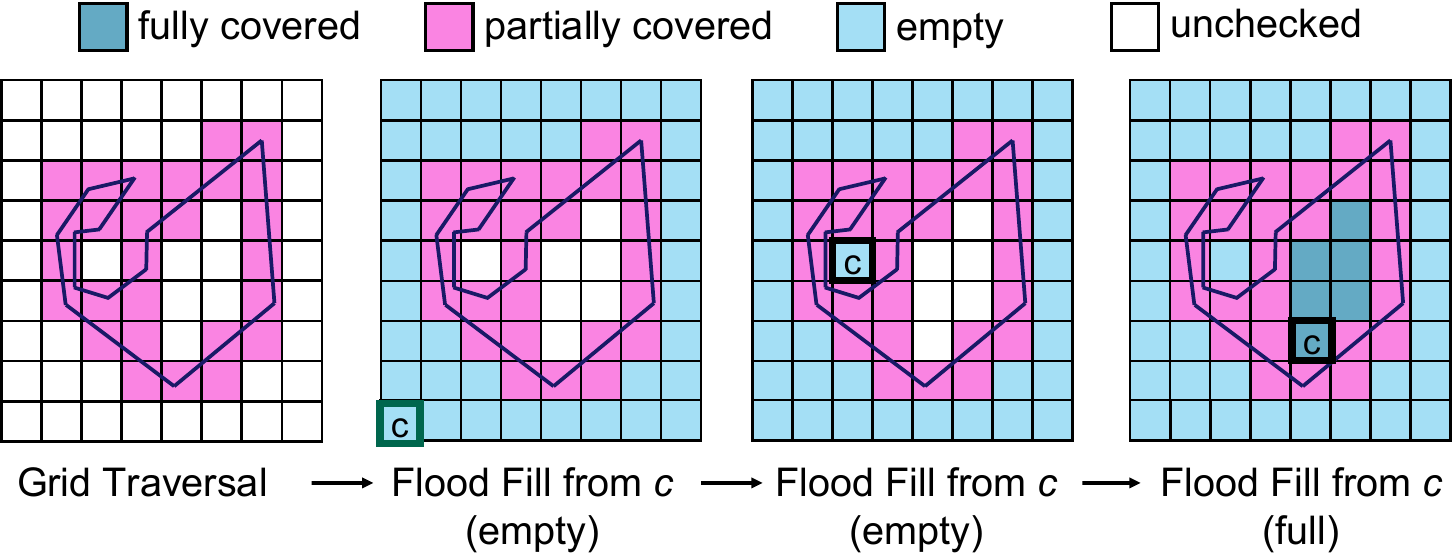}
	\vspace{-4ex}
	\caption{The flood fill algorithm, performing 3 iterations with different seeds \textit{c} to completely fill all unchecked cells.}	
	\label{fig:rasterization}
\end{figure}

Identifying the Partial cells is closely related to the pixel drawing problem in graphics that involves detecting which cells to ``turn on'' to draw a target line. While Bresenham's algorithm \cite{Bresenham65} is a popular and fast pixel drawing algorithm,
it approximates a line segment by turning on a minimal amount of cells and may thus not detect all intersected cells. 
In contrast, the Digital Differential Analyzer (DDA) method \cite{Museth14} is slower,
but identifies correctly and completely all intersected cells.
To detect the Partial cells, we use an efficient variant of DDA
\cite{AmanatidesW87} that uses grid traversal. 
We execute the grid traversal for each edge of the polygon and store the IDs of the identified Partial cells in a list.
The leftmost grid in Figure \ref{fig:rasterization} shows the Partial cells detected by the grid traversal algorithm for the polygon drawn in the figure.

\revj{Next, to identify the Full cells, we have to detect all cells that reside in the polygon's enclosed areas. This is closely related to the shape filling problem in computer graphics, a very old and thoroughly studied problem. Most approaches are variants of two paradigms: sweep-line (scanline) rendering \cite{CompGeomDBergMark} or shape (flood) filling \cite{Reynolds77,Smith79}. Both approaches have their pros and cons in terms of performance and accuracy.
	
\stitle{Scanline} Sweep-line algorithms use horizontal lines to find all intersections between them and the polygon per row in the grid, and sort the intersections based on their $x$-values. These are called \textit{event} points and are used to loop through all internal cells without performing any point-in-polygon (PiP) tests. All cells in-between consecutive pairs of the sorted event points are simply looped through and are labeled as Full. Note that this approach can be used as standalone method without the grid traversal algorithm, to detect all cells that overlap with a polygon regardless of type. However, in order to accurately classify each cell as Partial or Full and without performing a PiP test for every single one, we use it only for the Full cells, right after the DDA algorithm has generated the set of Partial cells.

\stitle{Flood Fill} The classic flood fill algorithm first selects an unlabeled cell that is guaranteed to be within the polygon, called \textit{seed}. Then, it traverses all neighboring cells of the seed until it finds the boundaries of the enclosed area, classifying the encountered cells as fully covered.
We implemented a variant of this algorithm which minimizes the number of  PiP  tests required to identify whether a cell is inside or outside the polygon.
Specifically,
we iterate through the cells of the polygon's MBR area.
If a cell $c$ has not been labeled yet (e.g., as Partial), we perform a PiP check from $c$'s center.
If the cell $c$ is found to be inside the polygon, $c$ is marked as Full and we perform a flood fill using $c$ as the seed, stopping at labeled cells,  and label all encountered unchecked cells as Full.
If the cell $c$ is found to be outside the polygon, $c$ is marked as Empty and we perform flood fill to mark Empty cells. The algorithm repeats as long as there are unchecked cells to flood fill from.
This reduces the number of PiP tests that need to be performed, as it suffices to perform a single test for each contiguous region in the grid with Full or Empty cells.
}

\begin{figure}[htb]
	\centering
	\includegraphics[width=\columnwidth]{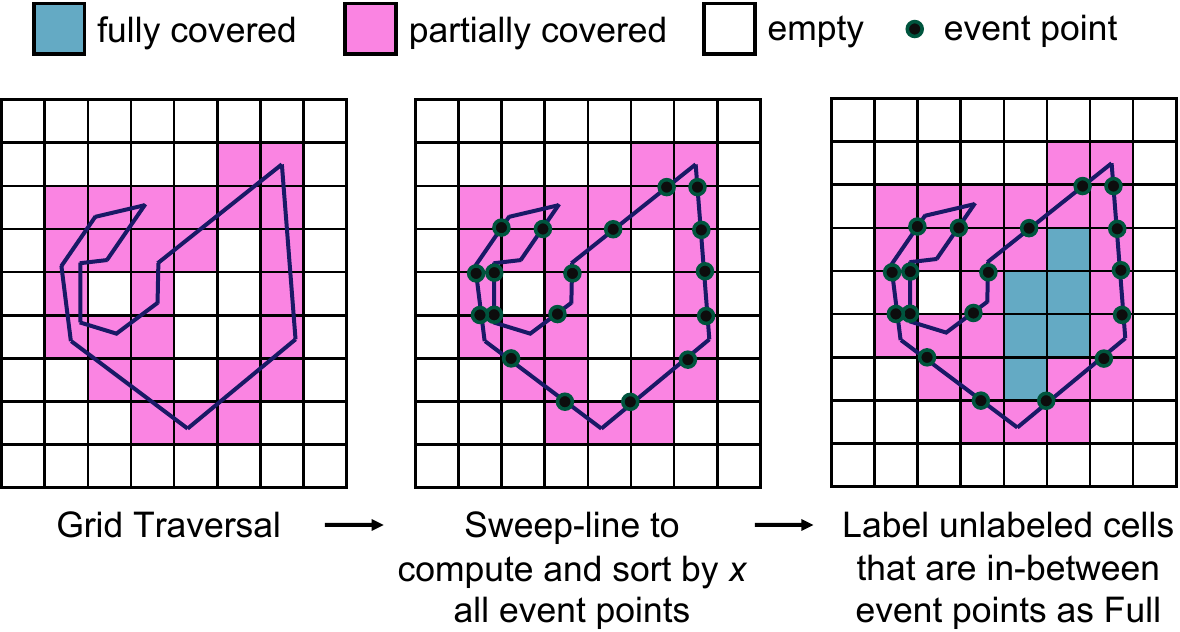}
	\vspace{-4ex}
	\caption{\revj{The Scanline rendering algorithm we implemented, filling the Full cells without performing any PiP tests.}}
	\label{fig:scanline}
\end{figure}

\revj{Figure \ref{fig:scanline} shows an example of our sweep-line rendering variant method. After all Partial cells have been identified, the event points between the scanlines and the polygon's edges are calculated and sorted by their $x$-values. Then, for each two consecutive pairs of event points, their in-between unlabeled cells are labeled as Full. 
}
Figure \ref{fig:rasterization} illustrates our Flood Fill process for an example polygon.
The unchecked cells form three contiguous regions bounded by Partial cells,
two of them being outside the polygon and one inside. 
Instead of looking for cells within the polygon to flood fill starting from them, it is faster to fill both the inside and outside of the polygon (marking cells as Full and Empty, respectively), as the number of point-in-polygon tests is minimized.

\revj{Regardless of which rasterization approach was chosen and after} all Partial and Full cells have been identified, the algorithm merges %\thanasis{Finally, we merge
consecutive cell identifiers into intervals to create the \textit{A}- and \textit{F}-lists that form the \methodname{} approximation.

\subsection{One-Step Intervalization}\label{subsec:onesteprasterization}
The approach described in the previous section identifies the types (Partial, Full, Empty) of all cells that intersect the MBR of the input polygon. For polygons which are relatively large and their MBRs define a large raster area this can be quite expensive.
We propose an alternative approach that identifies the {\em F}-intervals of the \methodname{} approximation efficiently and directly uses them to identify the {\em A}-intervals that include them in one step, without the need to identify the types of all individual cells in them.

\begin{figure}[htb]
	\centering
	\includegraphics[width=\columnwidth]{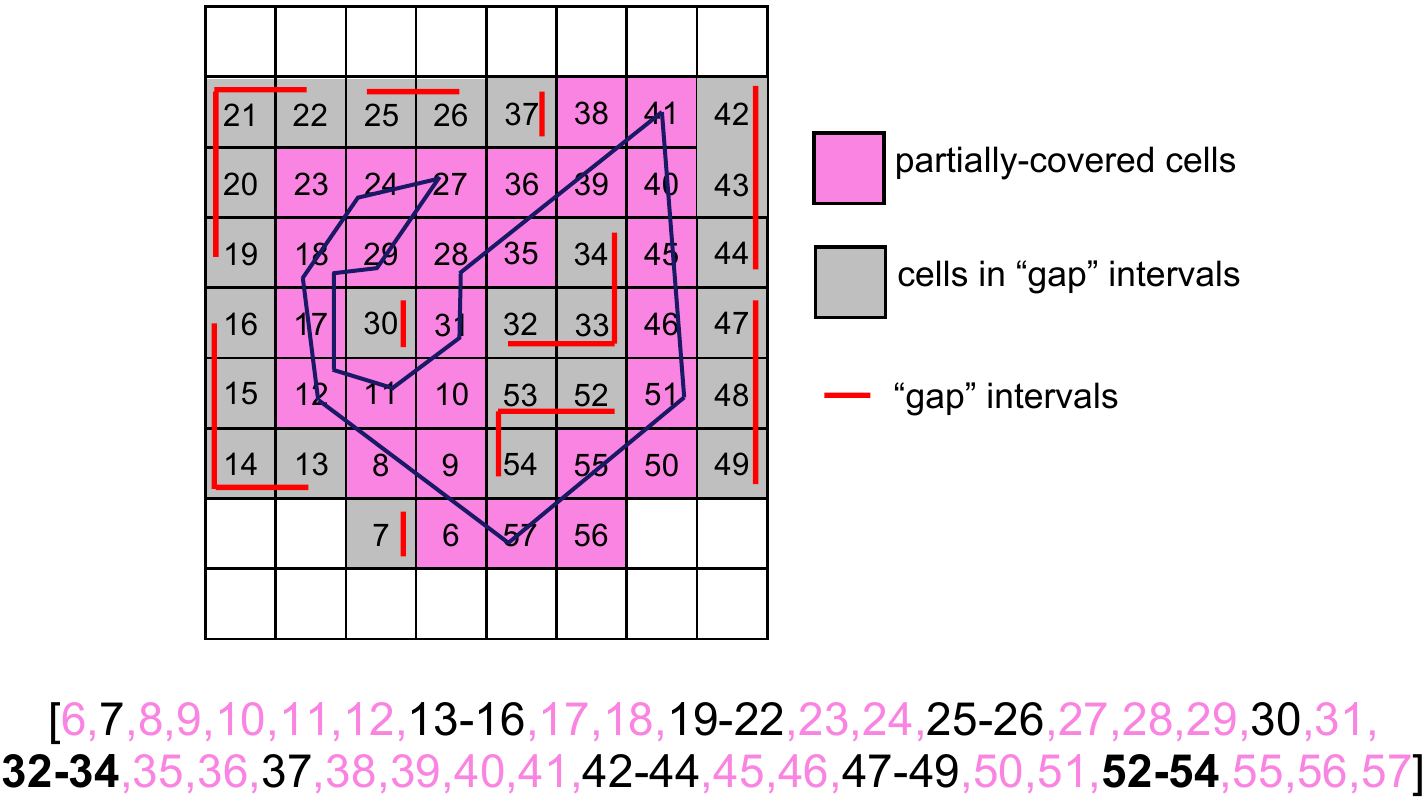}
	\vspace{-4ex}
	\caption{Example of the intervals/gaps for a set of Partial cells. Whether a gap will be labeled as Full or Empty, depends on the outcome of the PiP test.}	
	\label{fig:candidategaps}
\end{figure}

As in Sec. \ref{subsec:rasterization}, we first apply DDA
\cite{AmanatidesW87} to detect the Partial cells and sort them in Hilbert order.
An important observation is that ``gaps'' between nonconsecutive identifiers in the sorted Partial cells list, indicate candidate Full intervals on the Hilbert curve. Fig. \ref{fig:candidategaps} shows how these gaps are formed for an example polygon. Identifying the first cell $c$ of each candidate interval as Full or Empty, through a point-in-polygon (PiP) test, is enough to label the whole interval as Full or Empty, respectively.
The first ``gap'' interval is $[7,8)$ containing just cell $7$, which can be marked empty after a PiP test. From all ``gap'' intervals those marked in {\bf bold} (i.e., 32-34 = $[32,35)$ and 52-54 = $[52,55)$) are Full intervals and can be identified as such by a PiP test at their first cell (i.e., $32$ and $52$, respectively).  

Additionally, we can skip some of these PiP tests by checking all adjacent cells (north, south, west, east) of the first cell $c$ with smaller identifiers than $c$; if any of them is Full or Empty, we can also give the same label to the candidate interval, as it should exist in the same inner/outer area of the raster image. For example, in Figure \ref{fig:candidategaps}, when the algorithm moves to identify the interval $[52,55)$, it can detect that its first cell $52$ is adjacent to another Full cell with smaller order (cell $33$), that has been previously identified. Thus, the interval $[52,55)$ exists in the same inner area as cell $33$ and it inherits its label (Full), without performing another PiP test for it. In this example, a total of 5 PiP tests will be performed, for the intervals that start with the cells $7,13,30,32$ and $42$, instead of $11$ PiP tests that would be performed otherwise, if we did not take into consideration the neighboring cells. 

Algorithm \ref{alg:onestepalgorithm} is a pseudocode for the one-step intervalization process, which takes as input the sorted Partial cells list $P$ computed by DDA.
The algorithm creates the {\em A}-list, {\em F}-list
of the polygon in a single loop through $P$.
In a nutshell, the algorithm keeps track of the starting point of every {\em A}-interval and when an empty gap is identified, the algorithm ``closes'' the current {\em A}-interval and starts the next one from the next Partial cell in the list. On the other hand, Full intervals start with the identifier of the cell that is right after the last Partial cell of a consecutive sequence and end before the next Partial cell in order.

In details, Algorithm \ref{alg:onestepalgorithm}, starting from the first cell $p$ in $P$, keeps track of the starting cell-ID $Astart$ of the current {\em A}-interval; while the next cell $p+1$ in Hilbert order is also in $P$ (Lines \ref{lin:st:nextpartial}--\ref{lin:end:nextpartial}) the current {\em A}-interval is expanded. If the next cell $c=p+1$ is not partial, it is the starting cell of a candidate   {\em F}-interval. We first apply function $CheckNeighbors(c)$ to find whether there exists an adjacent cell of $c$ which is part of a $FULL$ or $EMPTY$ interval.
Specifically, for cell $c$ and a neighbor $n$, we first check whether $n < c$ (if not, $n$ is either Partial or unchecked); if yes, we binary-search $P$ to check whether $n$ is a $P$-cell. If not, we apply a special binary search method on
the current {\em F}-list to find out whether $n$ is part of an interval in it.
If we find $n$ as part of an {\em F}-interval, then $c$ is definitely a Full cell.
If we do no find $n$, then $c$ is definitely an Empty cell because $n<c$ and $n$ is not Partial.
If for all neighbors $n$ of $c$, either $n>c$ or $n$ is Partial, then we cannot determine the type of $c$ based on the current data, so we perform a  PiP test to determine $c$'s type (i.e., Full or Empty).
If $c$ is Full, then we know that the entire interval $[c,p)$ is  $FULL$ and append it to the  {\em F}-list (Line 16).
Otherwise ($c$ is Empty), $c$ is the end of the current {\em A}-interval, so the interval is added to the {\em A}-list and the start of the next {\em A}-interval is set to the next Partial cell $p$.
The algorithm continues until the list $P$ of partial cells is exhausted and commits the 
last {\em A}-interval (Line 23).

\begin{algorithm}
	\footnotesize
	\begin{algorithmic}[1]
		\Require{Sorted Partial cell array $P$}
		\Function{OneStepIntervalization}{$P$}

		\State $i\gets 0$ 	\Comment{current position in array $P$}
		\State $Astart\gets P_i$; $p \gets P_i$	\Comment{cell-IDs of current $A$-interval and partial cell} \label{lin:st:nextpartial}
		\While{$i< |P|$ and $p+1 = P_{i+1}$} \Comment{while next cell is partial}
			\State $i \gets i+1$
			\State $p \gets P_{i}$
		\EndWhile
		\State $c \gets p+1$	\Comment{next uncertain cell}
		\State $i \gets i+1$; $p \gets P_{i}$	\Comment{next partial cell}\label{lin:end:nextpartial}
		
		\While{$i< |P|$}
			
			\State $type \gets CheckNeighbors(c)$
			\If{$type \ne FULL$ and $type \ne EMPTY$} \Comment{$type$ is still uncertain}
                        \State $type \gets PointInPolygon(c)$ \Comment{PiP test gives $FULL$ or $EMPTY$}
                        \EndIf
			\If{$type = FULL$}
				\State $AppendFullInterval([c,p))$
			\Else \Comment{$type$ is $EMPTY$}		
				\State $AppendAllInterval([Astart,c))$ \Comment{current {\em A}-interval finalized}
				\State $Astart \gets p$ \Comment{start new {\em A}-interval}
			\EndIf
			\State Execute Lines  \ref{lin:st:nextpartial}--\ref{lin:end:nextpartial} \Comment{go through partial cells until next gap}
		\EndWhile 
		
		\State $AppendAllInterval([Astart,P_{i-1}+1))$	\Comment{save last $ALL$ interval}
		
		\EndFunction
	\end{algorithmic}
	\caption{The One-Step Intervalization algorithm.}
	\label{alg:onestepalgorithm}
\end{algorithm}

Our one-step intervalization approach performs $|P|-1$ PiP tests in the worst-case, which dominate its cost. 
Compared to the FloodFill-based approach of Section \ref{subsec:rasterization}, which explicitly marks and then sorts all Full and Partial cells,
Algorithm \ref{alg:onestepalgorithm} is expected to be much faster for polygons which are large compared to the cell size and include a huge number of Full cells. 
On the other hand, flood filling may be a better fit for small polygons with a small MBR and relatively few Full cells.

%% file: experiments.tex
\begin{table*}
	\centering
	\caption{Statistics of the datasets and space requirements of
		the data and the approximations}
	\vspace{-0.3cm}
	\label{tab:datasets}
	\sf
	
\resizebox{\textwidth}{!}{
		\begin{tabular}{|l|@{~}c@{~}|@{~}c@{~}|@{~} c@{~}|@{~}c@{~}|@{~} c@{~} |@{~} c@{~}|@{~}c@{~}|@{~} c@{~} |@{~} c@{~}|@{~}c@{~}|@{~} c@{~} |@{~}c@{~}|@{~}c@{~}|@{~} c@{~} |@{~} c|}
			\hline
			&T1&T2&T3&O5AF&O6AF&O5AS&O6AS&O5EU&O6EU&O5NA&O6NA&O5SA&O6SA&O5OC&O6OC\\\hline
			\# of Polygons&123K&2.25M&3.1K&72K&191K&447K&622K&1.9M&7.1M&4.0M&999K&123K&228K&107K&223K\\\hline
			Avg \# of vertices&25.4&31.9&2285.0&58.9&36.3&45.3&41.9&35.1&32.1&37.6&47.5&47.5&41.6&48.4&42.7\\\hline
			Avg obj MBR area&1.77E-04&4.03E-05&3.95E-01&2.03E-03&1.23E-03&1.03E-03&9.98E-04&1.25E-04&1.19E-04&1.11E-04&4.40E-04&1.34E-03&2.37E-03&5.00E-04&5.27E-04\\\hline \hline
			Geometries size (MB)&51.1&1168.1&115.3&68.9&112.7&327.9&422.1&1120.7&3746.2&2453.4&767.4&94.9&153.7&84.2&151.3\\\hline
			MBR size (MB)&4.4&81.1&0.1&2.6&6.9&16.1&22.4&70.9&258.4&144.8&36.0&4.5&8.2&3.9&8.1\\\hline
			APRIL size (MB)&14.4&134.0&57.2&14.2&25.4&55.2&64.5&180.3&968.0&251.0&155.0&25.4&44.4&7.3&15.0\\\hline
			APRIL-C size (MB)&6.6&75.3&16.0&5.1&10.6&23.3&28.6&84.8&406.5&138.0&62.4&9.2&16.7&3.8&7.8\\\hline
			RI size (MB)&19.5&138.2&968.7&18.6&55.7&57.5&109.8&180.9&942.9&238.1&213.5&31.2&143.4&14.2&39.3\\\hline
			RA size (MB)&1100.0&20000.0&26.9&617.2&1700.0&3700.0&5700.0&342.2&11400.0&6200.0&1500.0&1100.0&2100.0&898.7&2000.0\\\hline
			5C-CH size (MB)&28.7&705.4&1.6&18.5&46.6&117.8&159.4&515.4&1700.0&1200.0&257.7&30.4&52.9&28.8&57.7\\\hline
		\end{tabular}
	}
	\vspace{0.2cm}
\end{table*}

\section{Experimental Analysis}\label{sec:exp}

We assess the performance of our proposed methods (i.e., RI and APRIL), by
experimentally comparing them with previously proposed polygon
approximations for intermediate filtering of spatial joins. 
These include the
5-corner approximations comparison followed by a comparison of convex hulls 
(5C+CH) (as proposed in
\cite{BrinkhoffKSS94}),  and Raster Approximation (RA) of \cite{ZimbraoS98}.
We also included a baseline approach (None), which does not apply an
intermediate filter between the MBR-join and the refinement step.
For RA, we set the grid resolution to $K=750$ cells, except for a few
datasets where we use $K=100$, due to memory constraints.
For our methods (RI and APRIL), unless otherwise stated, we use a
granularity order $N = 16$ for the rasterization grid, meaning that
the Hilbert order of each cell can be represented by a 32-bit unsigned integer.
The MBR filter of the spatial join pipeline was implemented using the
algorithm of \cite{TsitsigkosBMT19}.
The refinement step was implemented using the Boost Geometry library
(www.boost.org) and its functions regarding shape intersection.
All code was written in C++ and compiled with the -O3 flag
on a machine with a 3.6GHz Intel i9-10850k and
32GB RAM, running Linux.

\subsection{Datasets}\label{subsec:datasets}

We used datasets from SpatialHadoop's \cite{SpatialHadoopDatasets}
collection.
T1, T2, and T3 represent
landmark, water and county areas in the United States (conterminous
states only).
We also used two Open Street Maps (OSM) datasets (O5 and O6) that
contain lakes and parks, respectively, from all around the globe.
We grouped objects into continents and created 6 smaller datasets
representing each one:
Africa (O5AF, O6AF), Asia (O5AS, O6AS), Europe (O5EU, O6EU), North
America (O5NA, O6NA), Oceania (O5OC, O6OC) and South America (O5SA,
O6SA).
From all datasets, we removed non-polygonal objects as well as
multi-polygons, self-intersecting polygons, \revj{and polygons with
  holes, because they need
  special handling by the refinement algorithms and the Boost Geometry
  library that we are using does not support some of these data
  types. The construction algorithms for APRIL approximations,
  presented in Section 6 can easily be adapted to handle these special
polygon classes with minor modifications; our raster
approximations for such polygons can directly be used as intermediate
filters.}
The first three rows of Table \ref{tab:datasets} show statistics about
the datasets.
The cardinalities of the datasets vary from 3.1K to 7.1M.
The smallest dataset (T3) includes complex polygons (thousands of
edges), each having a relatively large area (see third row of Table
\ref{tab:datasets}). The other datasets are larger and include medium
(e.g., T1, OSM data) to small and relatively simple polygons (e.g., T2).
We conducted spatial joins only between pairs of datasets that cover
the same area (i.e., T1 $\bowtie$ T2, T1 $\bowtie$ T3, O5AF $\bowtie$ O6AF, 
etc.).

\subsection{Optimizations and Customizations} \label{subsec:customization}

In this set of experiments, we showcase how the added features of \methodname{} perform both independently and  compared to RI. Additionally, we compare \methodname{} with RI in terms of space complexity, filter effectiveness, filter cost and creation time.

\subsubsection{The effect of $N$ in RI}\label{subsubsec:tuningRI}
Recall that our RI approach superimposes a $2^N\times 2^N$ grid over
the data space and approximates each object $o$ with the set $C_o$ of cells
that overlap with $o$. $C_o$ is then modeled by a set of intervals and
a bitstring for each interval, which encodes the types of the cells
that it contains.  
As discussed in Section \ref{sec:intervalizaton}, we set the value of
$N$ to 16, in order to have a fine granularity and be able to store
the interval endpoints in 4-byte unsigned integers. We confirm the appropriateness of this choice, by
evaluating the effectiveness of both RI and \methodname{} in spatial joins for
various values of $N$.

Table \ref{tab:rituning_time}
analyzes the performances of RI and \methodname{} for different values of $N$ in spatial
join T1~$\bowtie$~T2. The first three columns of the table show the percentage of candidate
pairs identified by the intermediate filters as true hits, false hits, or inconclusive (i.e.,
should be sent to the refinement step).
The last four columns show the cost of the filter step of the spatial
join (MBR-join), the total cost of applying our intermediate filters that use RI and \methodname{} to all
candidate pairs, the total cost of the refinement step, and the
overall join cost. 
The MBR-join cost is $N$-invariant, as this operation is independent
of the subsequent steps (intermediate filter, refinement). 
Observe that the number of
inconclusive pairs shrinks as $N$ increases; the
refinement cost decreases proportionally.
On the other hand, the cost of RI-filter increases with $N$ as the
intervals become more and longer.
Eventually, for the largest value of $N$, the overall join cost
converges to less than 1 second. 

In Table \ref{tab:rituning_stats}, 
we show the total time required to compute 
the RI and \methodname{} object approximations of all objects in T1
and T2
and the corresponding storage
requirements for them, as a function of $N$.
For small values of $N$, where the intermediate filters are not very effective, the computation
cost and the space requirements are low because, for each object, only a small number of
intervals, each approximating a small number of cells are constructed.
On the other hand, for large values of $N$, where the intermediate filters are most
effective, the approximations are very fine and require more time for
computation and more space.
We performed the same analysis for all other pairs of joined datasets
(results are not shown, due to space constraints)  and drew the same conclusions.
Overall, due to the high effectiveness for $N=16$, which
brings the best possible performance to the overall spatial join, we choose
this value of $N$ in the rest of the experiments.
\rev{Although we use a fixed grid for all objects (independently
	of their sizes), the intervalization and compression of the raster
	representations does not incur an unbearable space overhead and at
	the same time we achieve a very good filtering performance even for
	small objects, while avoiding re-scaling at runtime (as opposed to \cite{ZimbraoS98}).}

\begin{table}
	\centering
	\footnotesize
	\caption{Effect of $N$ on the performance of RI  and \methodname{} in T1$\bowtie$T2}
	\label{tab:rituning_time}
	\vspace{-0.3cm}
	\resizebox{\columnwidth}{!}{%
		\begin{tabular}{|l|@{~}c@{~}|@{~}c@{~}|@{~}c@{~}|@{~}c@{~}|@{~}c@{~}|@{~}c@{~}|@{~}c@{~}|}
			\hline
			&True hits&False hits&Indecisive&MBR-join (s)&RI-filter
			(s)&Refinement (s)&Total
			time(s)\\
			\hline
			&\multicolumn{7}{c|}{{\bf T1 $\bowtie$ T2 (RI)}}\\
			\hline
			$N=10$&5.68\%&24.96\%&69.36\%&0.03&0.03&1.44&1.50\\
			$N=13$&13.34\%&46.88\%&39.79\%&0.03&0.06&0.63&0.72\\
			$N=14$&17.74\%&52.20\%&30.06\%&0.03&0.09&0.48&0.60\\
			$N=15$&21.65\%&56.07\%&22.28\%&0.03&0.15&0.37&0.54\\
			$N=16$&24.50\%&59.42\%&16.08\%&0.03&0.28&0.27&0.59\\
			\hline
			&\multicolumn{7}{c|}{{\bf T1 $\bowtie$ T2 (APRIL)}}\\
			\hline
			$N=10$&5.67\%&24.96\%&69.37\%&0.03&0.03&1.45&1.52\\
			$N=13$&13.46\%&46.88\%&39.66\%&0.03&0.04&0.61&0.68\\
			$N=14$&17.99\%&52.20\%&29.81\%&0.03&0.04&0.45&0.52\\
			$N=15$&21.85\%&56.07\%&22.08\%&0.03&0.04&0.34&0.41\\
			$N=16$&24.29\%&59.42\%&16.29\%&0.03&0.05&0.26&0.34\\
			\hline
	\end{tabular} }
	%	\vspace{0.2cm}
\end{table}

\begin{table}
	\centering
	\footnotesize
	\caption{Effect of $N$ on the cost and space of RI and \methodname{} 
       for T1 and T2}
	\label{tab:rituning_stats}
	\vspace{-0.3cm}
	\resizebox{\columnwidth}{!}{%
		\begin{tabular}{|l|@{~}c@{~}|@{~}c@{~}|@{~}c@{~}|@{~}c@{~}|}
			\hline
			T1&RI constr. cost (s)&APRIL constr. cost (s)&RI Size (MB)&APRIL Size (MB)\\
			\hline
			$N=10$&0.98&0.29&2.6&3.0\\
			$N=13$&5.32&0.55&3.5&3.6\\
			$N=14$&13.90&0.85&4.7&4.4\\
			$N=15$&43.17&1.37&8.2&7.7\\
			$N=16$&148.72&2.37&19.0&13.8\\
			\hline
			T2&RI constr. cost (s)&APRIL constr. cost (s)&RI Size (MB)&APRIL Size (MB)\\
			$N=10$&15.29&5.68&46.0&53.0\\
			$N=13$&43.95&8.08&53.0&58.4\\
			$N=14$&87.35&11.23&62.0&66.7\\
			$N=15$&214.04&16.57&82.0&84.1\\
			$N=16$&620.57&26.76&132.0&128.0\\
			\hline
	\end{tabular} }
	%	\vspace{0.2cm}
\end{table}

\subsubsection{Join Order} \label{subsec:seqorderexperiments}

So far the interval joins in \methodname{} are assumed to be
applied in a fixed order: {\em AA}, {\em AF}, and {\em FA}.
As discussed in Section \ref{subsec:spatialjoins}, the joins can be
performed in any order.
Table \ref{tab:seqorder} tests different join orders for
T1 $\bowtie$ T2  and T1 $\bowtie$ T3.
T1 $\bowtie$ T2 (like the majority of tested joins) has a high
percentage of true negatives, so the original order is the most
efficient one (changing the order of  {\em AF} and {\em FA} does not
make a difference). On the other hand, for T1 $\bowtie$ T3, where the true
hits are more, pushing the {\em AA}-join at the end is more
beneficial.
Since knowing the number (or probability) of true negatives and true
hits a priori is impossible and because the join order does not make a
big difference in the efficiency of the filter (especially to the
end-to-end join time), we suggest using the fixed order, which is the
best one in most tested cases.  
In the future, we investigate the use of data statistics and/or object
MBRs to fast guess a good join order on an object pair basis.

\begin{table}
	\centering
	\footnotesize
	\caption{Join order effect on \methodname{} filter cost.}
	\vspace{-0.3cm}
	\label{tab:seqorder}
        \resizebox{\columnwidth}{!}{%
		\begin{tabular}{|l |@{~}c@{~}|@{~}c@{~}|@{~}c@{~}|@{~}c@{~}|}
			\hline
			{\bf Join Order}&{\bf True hits}&{\bf True negatives}&{\bf Indecisive}&{\bf Int. Filter (s)} \\
			\hline
			&\multicolumn{4}{c|}{{\bf T1 $\bowtie$ T2}}\\
			\hline
			{\bf AA-AF-FA}&24.29\%&59.42\%&16.29\%&0.0505\\
			{\bf AA-FA-AF}&24.29\%&59.42\%&16.29\%&{\bf 0.0501}\\
			{\bf AF-FA-AA}&24.29\%&59.42\%&16.29\%&0.0585\\
			{\bf FA-AF-AA}&24.29\%&59.42\%&16.29\%&0.0601\\
			\hline
			&\multicolumn{4}{c|}{{\bf T1 $\bowtie$ T3}}\\
			\hline
			{\bf AA-AF-FA}&69.84\%&28.13\%&2.03\%&0.1872\\
			{\bf AA-FA-AF}&69.84\%&28.13\%&2.03\%&0.1891\\
			{\bf AF-FA-AA}&69.84\%&28.13\%&2.03\%&{\bf 0.1737}\\
			{\bf FA-AF-AA}&69.84\%&28.13\%&2.03\%&0.1773\\
                  \hline
                \end{tabular} }
\end{table}

\subsubsection{Partitioning}\label{subsec:partitioningexperiments}
Tables \ref{tab:partitionsTime} and \ref{tab:partitionSizes}
illustrate the effect
of data partitioning
(Section \ref{subsec:partitioning})
on the effectiveness, query evaluation
time, and space requirements of \methodname{} approximations.
A higher number of partitions means finer-grained grids per partition
and thus, more intervals per polygon (i.e., more space is required).
Even though this reduces the amount of inconclusive cases,
it can slow down the intermediate filter, since more intervals need to
be traversed per candidate pair. For example, T1 $\bowtie$ T3 has
already a small percentage of inconclusive pairs, so partitioning may
not bring a significant reduction in the total join time.
On the other hand, for joins with high inconclusive percentage, such as
O5AS $\bowtie$ O6AS, partitioning can greatly reduce the total cost.
In summary, partitioning comes with a time/space tradeoff.

\begin{table*}
	\centering
	\caption{\# partitions per dimension effect on join time.}
	\label{tab:partitionsTime}
		\begin{tabular}{|l|@{~}c@{~}|@{~}c@{~}|@{~}c@{~}|@{~}c@{~}|}
			\hline
			{\bf \#}&{\bf Indecisive}&{\bf Int. Filter (s)}&{\bf Refinement (s)}&{\bf Total time (s)}\\
			\hline
			&\multicolumn{4}{c|}{{\bf T1 $\bowtie$ T2 }}\\
			\hline
			{\bf 1}&16.29\%&0.08&0.27&0.39\\
			{\bf 2}&12.81\%&0.06&0.22&0.32\\
			{\bf 3}&11.36\%&0.08&0.20&{\bf 0.30}\\
			{\bf 4}&10.50\%&0.09&0.20&0.32\\
			\hline	
			&\multicolumn{4}{c|}{{\bf T1 $\bowtie$ T3 }}\\
			\hline
			{\bf 1}&2.03\%&0.47&0.34&0.86\\
			{\bf 2}&1.77\%&0.29&0.29&{\bf 0.62}\\
			{\bf 3}&1.67\%&0.37&0.27&0.69\\
			{\bf 4}&1.64\%&0.49&0.26&0.80\\
			\hline	
			&\multicolumn{4}{c|}{{\bf O5AF $\bowtie$ O6AF }}\\
			\hline
			{\bf 1}&26.92\%&0.06&0.36&0.45\\
			{\bf 2}&21.24\%&0.06&0.29&0.37\\
			{\bf 3}&18.26\%&0.07&0.25&{\bf 0.34}\\
			{\bf 4}&16.63\%&0.08&0.24&0.35\\
			\hline	
			&\multicolumn{4}{c|}{{\bf O5AS $\bowtie$ O6AS }}\\
			\hline
			{\bf 1}&30.76\%&0.43&7.48&8.04\\
			{\bf 2}&24.07\%&0.41&5.30&5.83\\
			{\bf 3}&20.52\%&0.46&4.34&4.93\\
			{\bf 4}&18.39\%&0.55&3.61&{\bf 4.29}\\
			\hline	
			&\multicolumn{4}{c|}{{\bf O5EU $\bowtie$ O6EU }}\\
			\hline
			{\bf 1}&34.32\%&5.83&30.55&38.01\\
			{\bf 2}&27.97\%&5.35&24.24&31.22\\
			{\bf 3}&24.84\%&6.06&21.55&29.24\\
			{\bf 4}&22.60\%&6.61&19.99&{\bf 28.23}\\
			\hline	
			&\multicolumn{4}{c|}{{\bf O5NA $\bowtie$ O6NA }}\\
			\hline
			{\bf 1}&22.26\%&3.56&24.08&28.49\\
			{\bf 2}&17.58\%&3.14&18.81&22.81\\
			{\bf 3}&15.68\%&3.65&17.13&21.64\\
			{\bf 4}&14.45\%&4.52&16.02&{\bf 21.40}\\
			\hline	
			&\multicolumn{4}{c|}{{\bf O5SA $\bowtie$ O6SA }}\\
			\hline
			{\bf 1}&25.80\%&0.17&1.44&1.66\\
			{\bf 2}&20.74\%&0.14&1.21&1.39\\
			{\bf 3}&18.39\%&0.17&1.12&1.33\\
			{\bf 4}&17.03\%&0.20&1.07&{\bf 1.30}\\
			\hline	
			&\multicolumn{4}{c|}{{\bf O5OC $\bowtie$ O6OC }}\\
			\hline
			{\bf 1}&24.42\%&0.10&1.51&1.65\\
			{\bf 2}&18.89\%&0.12&1.09&1.25\\
			{\bf 3}&16.17\%&0.14&0.95&1.13\\
			{\bf 4}&14.65\%&0.16&0.88&{\bf 1.08}\\
			\hline	
		\end{tabular}
\end{table*}

\begin{table*}[htb]
	\centering
	\caption{\# of partitions per dimension, effect on \methodname{}
		size (MB).}
	\vspace{-0.3cm}
	\label{tab:partitionSizes}
        \resizebox{\textwidth}{!}{%
          \begin{tabular}{|@{~}l@{~}|@{~}c@{~}|@{~}c@{~}|@{~}c@{~}|@{~}c@{~}|@{~}c@{~}|@{~}c@{~}|@{~}c@{~}|@{~}c@{}|@{~}c@{}|@{~}c@{}|@{~}c@{}|@{~}c@{}|@{~}c@{}|@{~}c@{}|@{~}c@{}|}
			\hline
			\#&{\bf T1}&{\bf T2}&{\bf T3}&{\bf O5AF}&{\bf O6AF}&{\bf O5AS}&{\bf O6AS}&{\bf O5EU}&{\bf O6EU}&{\bf O5NA}&{\bf O6NA}&{\bf O5SA}&{\bf O6SA}&{\bf O5OC}&{\bf O6OC}\\\hline
			{\bf 1}&14.4&134.0&57.2&14.2&25.4&55.2&64.5&180.3&968.0&251.0&155.0&25.4&44.4&7.3&15.0\\
			{\bf 2}&26.1&236.3&112.0&29.2&49.2&106.9&124.2&336.9&1900.0&453.4&311.8&51.5&86&14.3&49.2\\
			{\bf 3}&37.1&352.6&166.7&44.7&74.2&164.0&188.3&492.5&2800.0&654.2&459.6&76.9&129.8&35.2&76.3\\
			{\bf 4}&47.2&465.9&224.9&61.4&99.5&219.1&255.1&653.0&3700.0&875.1&619.0&104.2&172.3&49.1&107.7\\
			\hline
		\end{tabular}
	}            
\end{table*}

\subsubsection{Different Granularity}\label{subsec:diffgranularityexperiments}

As discussed in Section \ref{subsec:differentGranularity}, we can 
define and use \methodname{} at lower granularity than $N=16$ for one or both
datasets, trading filter effectiveness for space savings.
In Table \ref{tab:diffgranularitytests}, we study the effect 
of reducing $N$ for T3 in T1 $\bowtie$ T3.
The size of T3's \methodname{} approximations halves every time we
decrease $N$ by one.
The filter time also decreases, due to the reduced amount
of intervals from T3 in the interval joins.
However, the percentage of indecisive pairs increases, raising the refinement cost.
$N=15$ is the best value for T3, because it achieves the same
performance as $N=16$, while cutting the space requirements in half.

\begin{table*}
	\centering
	\footnotesize
	\caption{Join between T1 (order 16) and T3 (order $N$).}
	\vspace{-0.3cm}
	\label{tab:diffgranularitytests}
		\begin{tabular}{|l |@{~}c@{~}|@{~}c@{~}|@{~}c@{~}| @{~}c@{~}|@{~}c@{~}|@{~}c@{~}|@{~}c@{~}|}
			\hline
			{\bf $N$}&{\bf True hits}&{\bf True negs.}&{\bf Indecisive}&{\bf Int. Filter (s)}&{\bf Refinement (s)}&{\bf Total (s)}&{\bf T3 size (MB)} \\
			\hline
			{\bf 16}&69.84\%&28.13\%&2.03\%&0.19&0.33&0.57&57.2\\
			{\bf 15}&69.63\%&27.85\%&2.52\%&0.13&0.41&0.59&28.3\\	
			{\bf 14}&69.18\%&27.46\%&3.36\%&0.11&0.54&0.70&14.0\\		
			{\bf 13}&68.39\%&26.86\%&4.75\%&0.09&0.78&0.92&6.9\\					
			{\bf 12}&66.63\%&25.70\%&7.67\%&0.09&1.23&1.37&3.4\\			
			\hline			
                \end{tabular}
\end{table*}

\subsection{\methodname{} Construction Cost} \label{subsec:rasterizationexperiments}

We now evaluate the
\methodname{} construction techniques that we have proposed
in Section \ref{sec:rasterization}, comparing them with the
rasterization method
used in previous work \cite{ZimbraoS98} (and for RI).
Note that RA \cite{ZimbraoS98} and RI
essentially apply 
polygon clipping and polygon-cell intersection area computations,
because they need to classify the cells that intersect the polygon
to Weak, Strong, and Full. On the other hand, \methodname{} uses two
classes: Partial and Full, which enables the application of the
techniques
that we proposed in Section \ref{sec:rasterization}.
Table \ref{tab:rasterizationTechniques} shows the
time taken to compute the \methodname{} approximations of all polygons
in each dataset (for $N=16$), using (i)  the 
rasterization+intervalization approach of RI,
after unifying Strong and Weak cells, (ii) the
\revj{Scanline and FloodFill approaches tailored for \methodname{}} presented in Section
\ref{subsec:rasterization}, and (iii) two versions of our novel OneStep
intervalization approach (Section
\ref{subsec:onesteprasterization}): one that performs a
point-in-polygon (PiP) test for each first cell $c$ of a candidate Full
interval and one that checks the Neighbors of $c$ before attempting
the PiP test. \revj{All costs in Table \ref{tab:rasterizationTechniques} include the intervalization cost as well, to generate the final interval lists needed for APRIL. The intervalization for the rasterization techniques is performed by merging cells with consecutive Hilbert order identifiers into intervals, while our methods from Section \ref{subsec:onesteprasterization} generate the intervals straight away.}

\begin{table}
	\centering
	\caption{\revj{Total construction cost (sec) for all datasets.}}
	\label{tab:rasterizationTechniques}
	\resizebox{\columnwidth}{!}{
		\begin{tabular}{|l|@{~}c@{~}|@{~}c@{~}|@{~} c@{~}|@{~}c@{~}|@{~}c@{~}|@{~}c@{~}|}
			\hline
			{\bf Dataset}&{\bf RI}&{\bf Flood Fill}&{\bf Scanline}&{\bf IDEAL}&{\bf OneStep (PiPs)}&{\bf OneStep (Neighbors)}\\\hline
			{\bf T1}&143.62&3.90&3.63&9.76&3.74&{\bf 2.19}\\\hline
			{\bf T2}&601.67&28.05&{\bf 23.06}&37.40&33.76&23.43\\\hline
			{\bf T3}&9919.06&265.72&278.50&666.19&75.40&{\bf 28.33}\\\hline
			{\bf O5AF}&264.45&4.25&{\bf 3.98}&13.02&11.00&4.72\\\hline
			{\bf O6AF}&468.47&13.06&12.62&32.45&5.66&{\bf 4.17}\\\hline
			{\bf O5AS}&486.86&11.69&{\bf 10.07}&27.42&21.28&11.78\\\hline
			{\bf O6AS}&994.93&28.98&{\bf 24.76}&56.14&65.01&25.07\\\hline
			{\bf O5EU}&1193.71&36.08&{\bf 30.30}&58.18&55.79&33.71\\\hline
			{\bf O6EU}&5493.15&172.20&{\bf 147.29}&244.95&243.17&156.94\\\hline
			{\bf O5NA}&1530.92&53.33&{\bf 45.26}&72.34&133.39&66.60\\\hline
			{\bf O6NA}&1630.29&43.40&40.89&76.62&51.79&{\bf 30.71}\\\hline
			{\bf O5SA}&361.87&6.67&{\bf 5.79}&20.69&14.74&6.77\\\hline
			{\bf O6SA}&1478.05&34.56&34.15&98.20&22.86&{\bf 10.52}\\\hline
			{\bf O5OC}&39.99&2.88&{\bf 2.48}&6.17&3.82&2.49\\\hline
			{\bf O6OC}&113.99&9.32&{\bf 8.49}&18.29&20.75&8.56\\\hline
	\end{tabular}}
\end{table}

\revj{We also included in the comparison the rasterization technique
  proposed for IDEAL \cite{Teng21}, which also detects Full and
  Partial cells, as implemented in \cite{IDEALimplementation}. We
modified IDEAL's granularity definition formula accordingly to match
\methodname{}'s Hilbert space grid of order $N=16$.} 

\revj{Observe that our OneStep intervalization algorithm employing the
Neighbors check (Section 6.2) and our Scanline rendering (Section 6.1)
are the fastest approaches in most of the cases. Scanline and Flood Fill show little to no difference in performance between them, with Scanline being overall faster for small polygons and Flood Fill being faster for large and more complex shapes. This is due to the fact that Flood Fill performs some PiP tests while Scanline does not, as well as Flood Fill filling some unecessary (Empty) pixels outside of the polygon, while Scanline focusing entirely on the interior of the shape specified by the event points}.
OneStep (Neighbors) applies $40\%-70\%$ fewer PiP tests compared to
OneStep (PiPs) that does not apply the 
Neighbors check.
\revj{Only in datasets containing
relatively small polygons
OneStep (Neighbors) is up
to 32\% slower than the Scanline method, however in most such cases their difference is negligible.
On the other hand, in some datasets containing large polygons (e.g., T3, O6AF,
O6SA)
OneStep is up to one order of magnitude faster than Scanline (T3) and
33\% to 224\% faster than the rest of the methods.
All methods proposed in Section \ref{sec:rasterization} are 
orders of magnitude faster compared to rasterization for RI, because
the latter has to perform expensive detection for Strong and Weak cells.}

\begin{figure*}[htb]
	\centering
				\begin{tabular}{@{}r@{~}r r@{~}r@{}}
					\includegraphics[width=0.42\columnwidth]{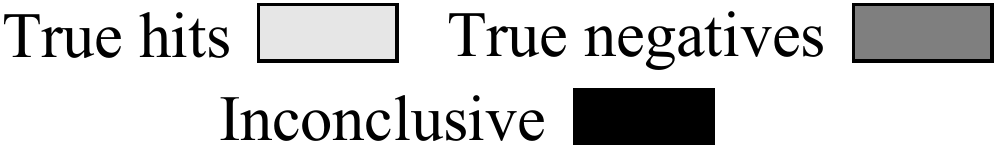}&
					\includegraphics[width=0.45\columnwidth]{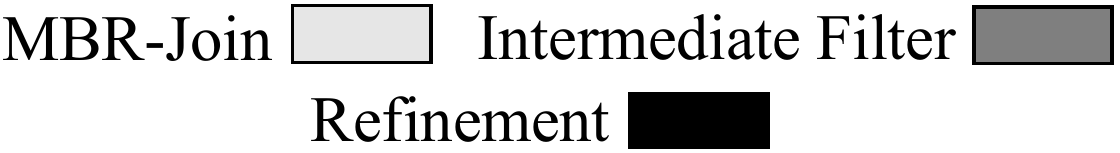}&
					\includegraphics[width=0.42\columnwidth]{plot_legend_effectiveness3}&
					\includegraphics[width=0.45\columnwidth]{plot_legend_cost3}\\
					\includegraphics[width=0.49\columnwidth]{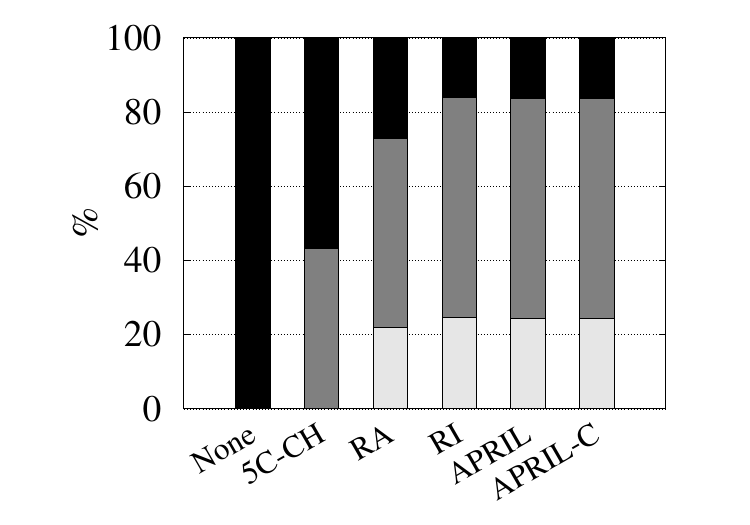}&
					\includegraphics[width=0.49\columnwidth]{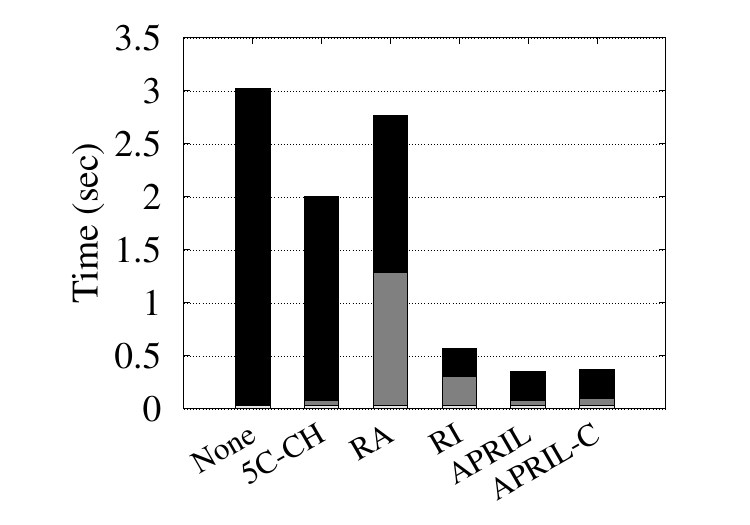}&
					\includegraphics[width=0.49\columnwidth]{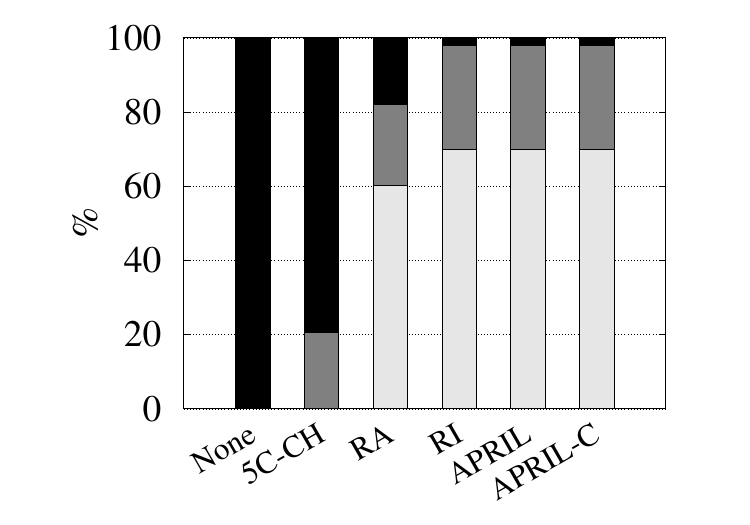}&
					\includegraphics[width=0.49\columnwidth]{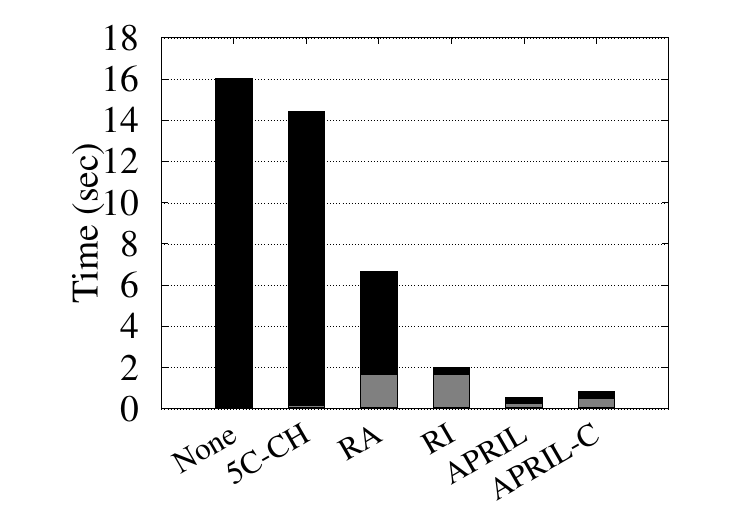}\\
					\multicolumn{2}{c}{(a) T1 $\bowtie$ T2}&
					\multicolumn{2}{c}{(b) T1 $\bowtie$ T3}\\ 
					
					\includegraphics[width=0.49\columnwidth]{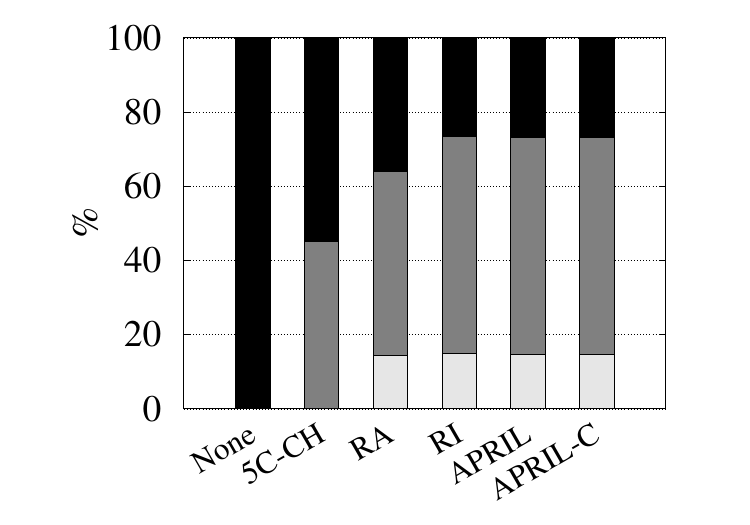}&
					\includegraphics[width=0.49\columnwidth]{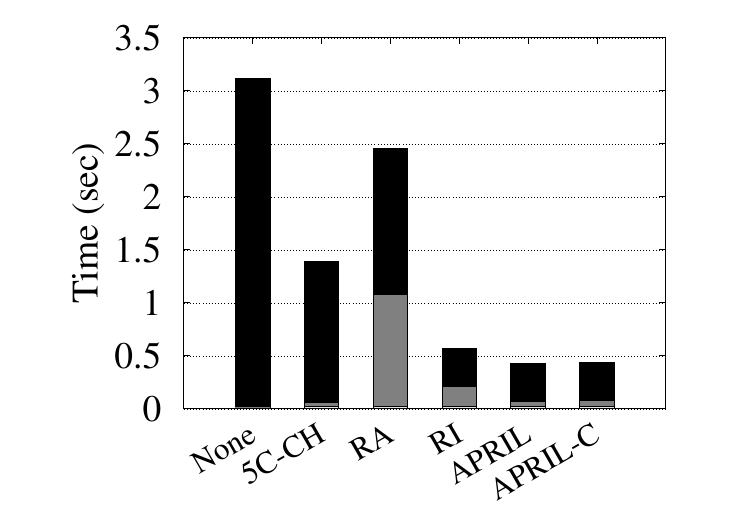}&
					\includegraphics[width=0.49\columnwidth]{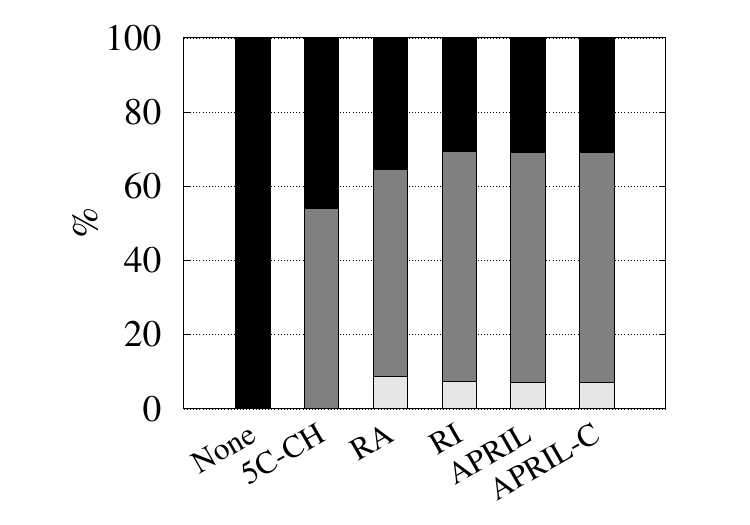}&
					\includegraphics[width=0.49\columnwidth]{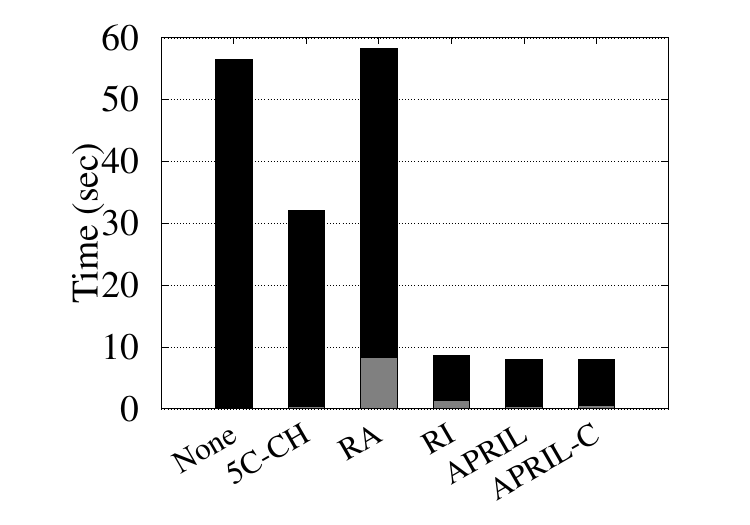}\\
					\multicolumn{2}{c}{(a) O5AF $\bowtie$ O6AF}&
					\multicolumn{2}{c}{(b) O5AS $\bowtie$ O6AS}\\ 
					
					\includegraphics[width=0.49\columnwidth]{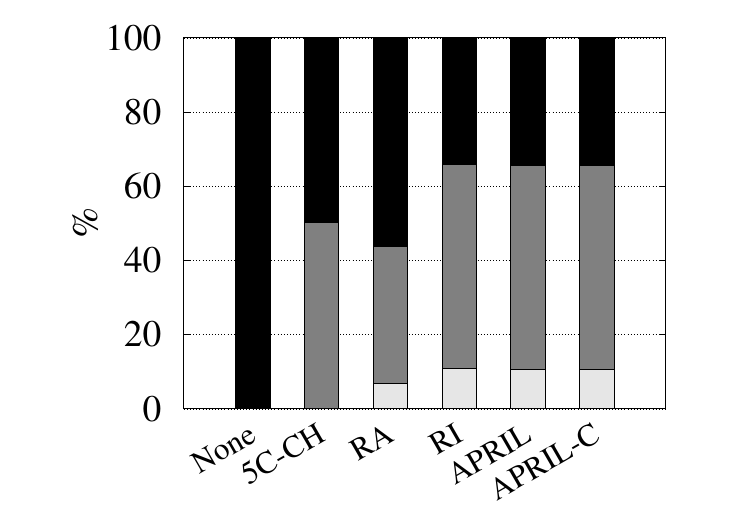}&
					\includegraphics[width=0.49\columnwidth]{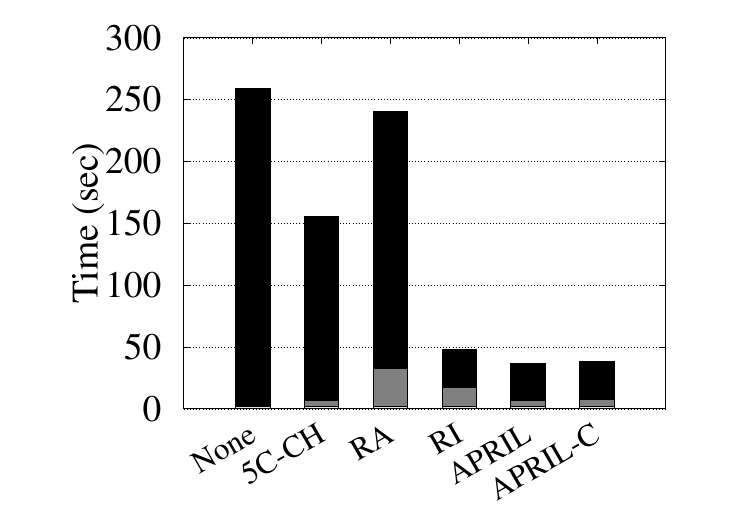}&
					\includegraphics[width=0.49\columnwidth]{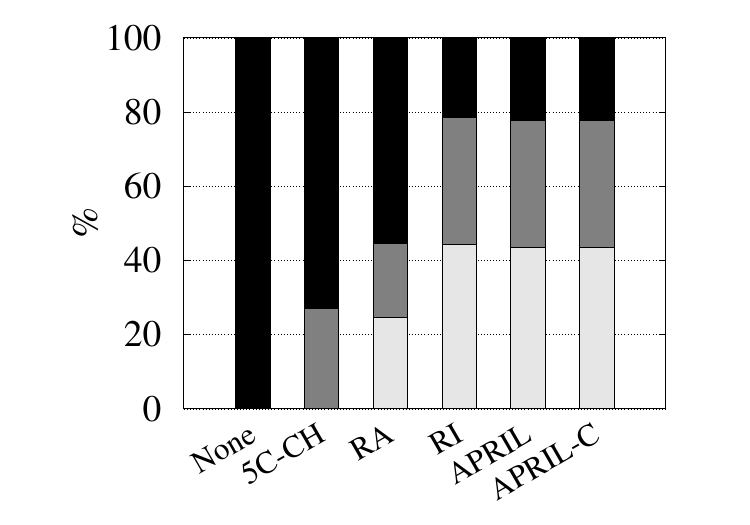}&
					\includegraphics[width=0.49\columnwidth]{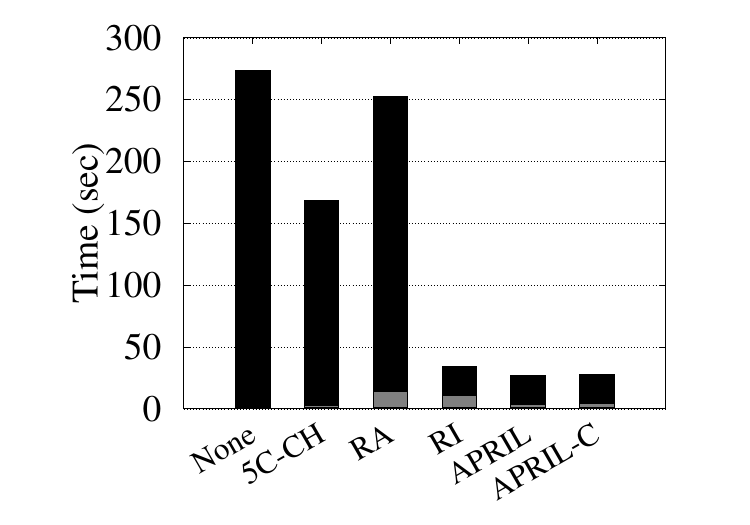}\\
					\multicolumn{2}{c}{(a) O5EU $\bowtie$ O6EU}&
					\multicolumn{2}{c}{(b) O5NA $\bowtie$ O6NA}\\ 
					
					\includegraphics[width=0.49\columnwidth]{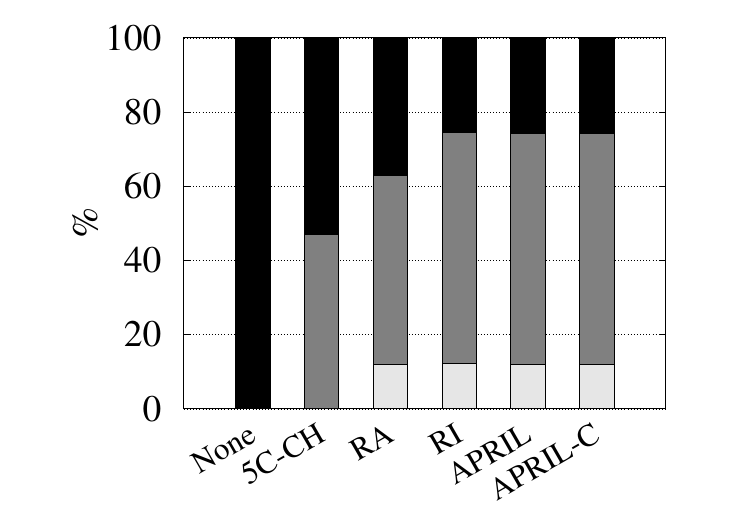}&
					\includegraphics[width=0.49\columnwidth]{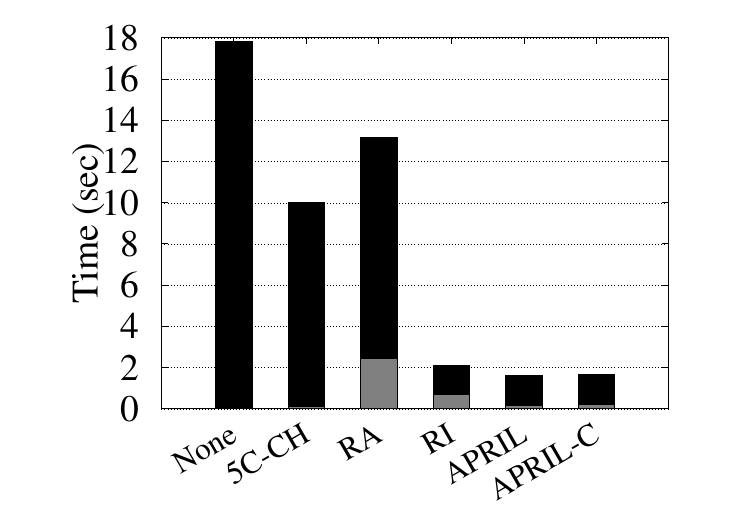}&
					\includegraphics[width=0.49\columnwidth]{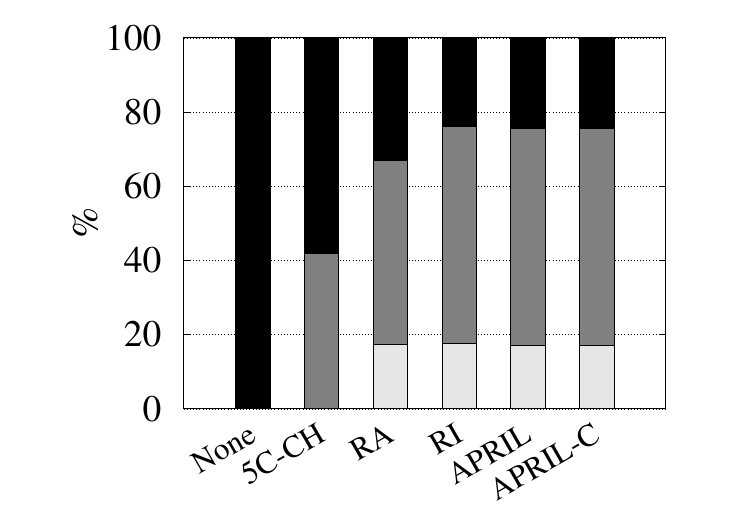}&
					\includegraphics[width=0.49\columnwidth]{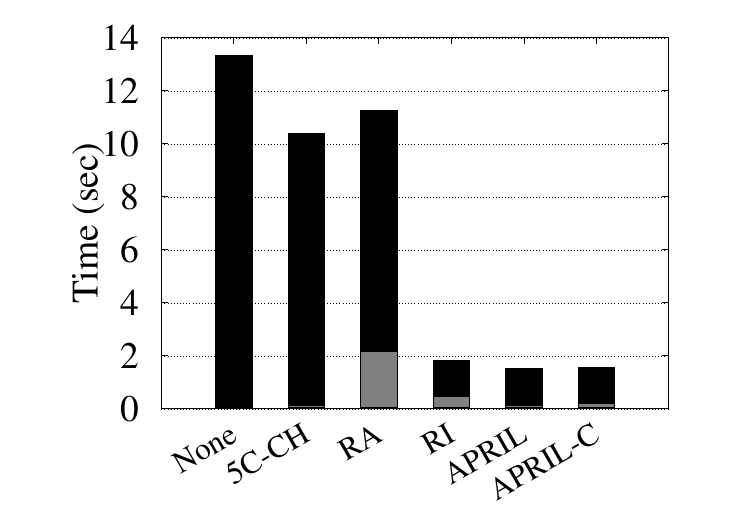}\\
					\multicolumn{2}{c}{(a) O5SA $\bowtie$ O6SA}&
					\multicolumn{2}{c}{(b) O5OC $\bowtie$ O6OC}\\ 
					
				\end{tabular}
				\caption{Filter effectiveness and spatial join cost for various
					intermediate filters.}	
				\label{fig:effectiveness}
\end{figure*}

\subsection{Comparative Study}\label{subsec:comparativestudy}

Finally, we compare RI and \methodname{} with other
intermediate filters in terms of space complexity, filter
effectiveness, and filter cost. 
For all experiments, we created \oldmethod{} and \methodname{} 
using a single partition
(i.e., the map of the two datasets that are joined in each case),
rasterized on a $2^{16} \times 2^{16}$ grid,
which is the best performing granularity for both
methods.
We used a fixed order ({\em AA}-, {\em AF}-, {\em FA}-) for the
interval joins of \methodname{}, as shown in Algorithm \ref{alg:apriljoin}.

\subsubsection{Space Complexity}\label{subsec:space}

Table \ref{tab:datasets} shows the total space requirements of the object
approximations required by each intermediate filter, for each of the datasets
used in our experiments.
\methodname{}
and \methodname{}-C refer to the uncompressed and compressed version
of \methodname{}, respectively. 
As a basis of comparison we also show the total space required to
store the exact geometries of the objects and their MBRs.
Note that, in most cases, our methods (RI, \methodname{} and \methodname{}-C) are
significantly more space efficient compared to RA and have similar or
lower space requirements to the 5C-CH. The only exception is T3, which
includes huge polygons that are relatively expensive to approximate
even by \methodname{}-C.
Notably, for most datasets, the
compressed \methodname{} approximations have similar space requirements as the object MBRs, meaning that we can keep them in memory and use them in main-memory spatial joins \cite{NobariQJ17} directly after the MBR-join step, without incurring any I/O.

\subsubsection{Comparison in Spatial Intersection Joins}\label{subsec:spatialjoinsexperiments}

We evaluate \methodname{} (both compressed and uncompressed version),
5C+CH, RA, and \oldmethod{}, on all join pairs, in
Figure \ref{fig:effectiveness}.
We compare their ability to detect true hits and true
negatives,
their computational costs as filters, and their impact to the
end-to-end cost of the spatial join.

\stitle{Filter Effectiveness}
\methodname{} and \oldmethod{} have the highest filter effectiveness among all
approximations across the board.
\methodname{}'s true hit ratio is slightly smaller compared to that of
RI because \methodname{} fails to detect the (rare) pairs of polygons
which only have Strong-Strong common cells.
However, this only brings a marginal
increase in the refinement step's cost, at the benefit of having a faster and
more space-efficient filter.
In O5AS $\bowtie$ O6AS and O5OC $\bowtie$
O6OC, \methodname{} and \oldmethod{}  have marginally lower true hit ratio compared to RA;
however, in these cases their true negative ratio is much higher than that of RA.
The least effective filter is 5C+CH, mainly due to its inability to detect true hits.

\stitle{Intermediate Filter cost}
5C+CH are simple approximations (a few points each), therefore the
corresponding filter is very fast to apply.
Notably, \methodname{} has a filtering cost very close to that of
5C+CH and sometimes even lower.
This is due to \methodname{}'s ability to model a raster
approximation as two sequences of integers, which are processed
by a sequence of efficient merge-join algorithms.
5C+CH has poor filtering performance,
which negatively affects
the total
join cost (last column), whereas
\methodname{} is very fast and very
effective at the same time.
The state-of-the-art filter \oldmethod{}
is more
expensive than \methodname{},
because it requires the
alignment and bitwise $AND$ing of the interval bit-codes.
As a result, \methodname{} is 3.5-8.5 times faster as an
intermediate filter compared to \oldmethod{}
(note the ``Intermediate Filter'' part of the cost in the bars). 
A comparison between the filter costs of \methodname{} and
\methodname{}-C reveals that decompressing the interval lists while
performing the joins in \methodname{}-C only brings a small overhead,
making compression well worthy, considering the space savings it
offers (see Table \ref{tab:datasets}).
The decompression cost is significant only in T1 $\bowtie$ T3, because
T3's \textit{A}-lists and \textit{F}-lists are quite long. Still, even
in this case, \methodname{}-C is much faster than \oldmethod{}.
\stitle{Refinement cost}
The refinement cost is intertwined with the 
percentage if indecisive pairs.
The detection of fewer candidate pairs as true hits or
true negatives leads to a higher refinement workload; this is why
\methodname{} and RI result in the lowest refinement cost, compared to
the rest of the approximations.

\stitle{Overall join cost}
\methodname{} (Section \ref{sec:methodology})
reduces the overall cost of end-to-end spatial
joins up to 3.5 times compared to using our 
\oldmethod{} intermediate filter (Section \ref{sec:ri}),
while also achieving a speedup of 3.23x-25x against the
rest of the approximations. Adding the \methodname{} intermediate
filter between the MBR filter and the refinement step
reduces the spatial join cost by 7x-28x.
\methodname{}'s high filtering effectiveness, low application
cost, and low memory requirements render it a superior 
approximation for filtering pairs in spatial intersection join pipelines.

\revj{
\stitle{\methodname{} vs. \oldmethod{}}
In summary, APRIL prevails over \oldmethod{} on all aspects
including space complexity, construction time, 
filtering efficiency, and overall spatial intersection join time.
Table \ref{tab:perfAPRIL}, summarizes the improvement that 
APRIL achieves over RI on all join pairs.

\begin{table}
	\centering
	\caption{Improvement of APRIL over RI on all join pairs}
	\label{tab:perfAPRIL}
        \resizebox{\columnwidth}{!}{
          \begin{tabular}{|c|c|c|}
  \hline
  {\bf APRIL vs. RI}& {\bf range} & {\bf average}\\\hline \hline
APRIL size (times smaller)&0.95x--16.94x&2.59x\\\hline
APRIL-C size (times smaller)&1.73x--60.54x&7.39x\\\hline
Construction (times faster)&13.32x--350.13x&70.71x\\\hline
Intermediate Filter (times faster)&3.45x--8.56x&4.86x\\\hline
End-to-end Join (times faster)&1.10x--3.51x&1.58x\\\hline
\end{tabular}}
\end{table}

}

\revj{

  \begin{table*}
	\centering
	\caption{End-to-end join performance between T2 and datasets
          with varying average object area.}
	\label{tab:variance}
		\begin{tabular}{|c|c|c|c|c|c|c|}
			\hline
		{\bf Method}&{\bf Accepted}&{\bf Rejected}&{\bf Inconclusive}&{\bf Int. Filter (s)}&{\bf Refinement (s)}&{\bf Total time (s)} \\\hline

                        \multicolumn{7}{c|}{{\bf T2 $\bowtie$ T1 }}\\
			\hline
		None&0.00\%&0.00\%&100.00\%&0.00&2.94&2.98 \\
		RA (K=750)&21.98\%&50.76\%&27.26\%&1.16&1.20&2.40 \\
		5C+CH&0.00\%&43.15\%&56.85\%&0.05&1.65&1.74 \\
                  APRIL&24.29\%&59.42\%&16.29\%&0.05&0.27&0.35 \\
                  \hline
		\multicolumn{7}{c|}{{\bf T2 $\bowtie$ T10 }}\\
			\hline
		None&0.00\%&0.00\%&100.00\%&0.00&296.01&297.51 \\
		RA (K=750, K=150)&27.49\%&25.15\%&47.36\%&27.61&228.51&257.55 \\
		5C+CH&0.00\%&29.11\%&70.89\%&2.28&204.66&208.68 \\
		APRIL&51.92\%&45.59\%&2.50\%&2.32&6.46&9.51 \\                  
			\hline	
			\multicolumn{7}{c|}{{\bf T2 $\bowtie$ T3 }}\\
			\hline
		None&0.00\%&0.00\%&100.00\%&0.00&310.81&312.38 \\
		RA (K=750, K=150)&58.27\%&23.15\%&18.58\%&31.82&102.67&135.79 \\
		5C+CH&0.00\%&22.03\%&77.97\%&2.32&224.14&228.40 \\
		APRIL&68.47\%&29.88\%&1.64\%&3.22&5.29&9.32 \\
			\hline	
			\multicolumn{7}{c|}{{\bf T2 $\bowtie$ T9 }}\\
			\hline
		None&0.00\%&0.00\%&100.00\%&0.00&2595.44&2596.96 \\
		RA (K=750, K=150)&49.34\%&20.26\%&30.40\%&24.55&1352.44&1378.21 \\
		5C+CH&0.00\%&21.70\%&78.30\%&2.36&2003.41&2007.63 \\
		APRIL&68.04\%&31.80\%&0.16\%&20.52&3.03&24.46 \\
                  \hline	
		\end{tabular}
              \end{table*}
              
\subsubsection{Effect of variance in object sizes}\label{subsec:variance}
We now test the effect that the variance between the sizes of joined
objects has on the performance of APRIL compared to previous work. For
this, besides T1, T2, and T3, we used two more Tiger datasets, i.e.,
T9 (States) and T10 (Zip codes). Table \ref{tab:sizesTIGER} lists
statistics of all
the Tiger datasets that we used in this experiment ordered by average area of the objects in them.  

To demonstrate APRIL's performance on dataset-pairs of varying object sizes, we joined the Water Areas dataset (T2), which has polygons with the smallest areas on average compared to other  Tiger datasets, in increasing order of average area per object.
Note that as the objects which are joined with T2-objects
grow larger (T1, T10, T3 and then T9) the indecisive cases reduce
drastically.
This can be explained by the fact that as a polygon grows larger, it generates more Full intervals and thus, the probability of detecting
a true hit between it and another polygon using APRIL increases.
Table \ref{tab:variance}
shows a detailed performance comparison for the spatial intersection join between the TIGER datasets. In all cases, APRIL retains the best filtering effectiveness and total execution time.
The performance gap between APRIL and the other methods grows
with the difference in sizes between the objects in the candidate pairs,
due to APRIL's effectiveness in detecting true hits, avoiding their costly refinement.

\begin{table}
  \centering
  \scriptsize
	\begin{tabular}{@{}|@{~}c@{~}|@{~}c@{~}|@{~}c@{~}|@{~}c@{~}|@{}}
		\hline
		{\bf Dataset}&{\bf \# objects}&{\bf avg \#
                                                vertices}&{\bf avg
                                                           MBR area }\\\hline
		T2 (Water areas)&2252316&31.9&4.03E-05 \\\hline
		T1 (Landmarks)&123045&25.4&1.77E-04 \\\hline
		T10 (Zip codes)&26091&1404.8&4.14E-02 \\\hline
		T3 (Counties)&3043&2316.2&3.95E-01 \\\hline
		T9 (States)&43&18140.4&2.59E+01 \\\hline
	\end{tabular}
	\caption{TIGER dataset statistics, sorted by ascending average object MBR area.}	
	\label{tab:sizesTIGER}
\end{table}

}

\subsubsection{Performance in other queries}\label{subsec:otherqueriesexperiments}
We now evaluate the performance of \methodname{}
in other queries, besides spatial
intersection joins.
We start with selection
queries of arbitrary shape (see Section \ref{subsec:selection}).
For this experiment, we sampled 1000
polygons from T3 and applied them as selection queries
on T1 and T2, simulating queries of the form: find all landmark
areas (T1) or water areas (T2) that intersect with a given US county
(T3).
As Table \ref{tab:competitorsrange} shows, compared to \oldmethod{},
\methodname{}
achieves a 3.5x-4x speedup in the total query cost.

\begin{table*}
	\centering 
	\footnotesize
	\caption{\methodname{} vs. \oldmethod{} (polygonal range queries).}
	\label{tab:competitorsrange}
	\begin{tabular}{|l |@{~}c@{~}|@{~}c@{~}|@{~}c@{~}|@{~}c@{~}| @{~}c@{~}|@{~}c@{~}|}
		\hline
		&{\bf True hits}&{\bf True negatives}&{\bf Indecisive}&{\bf Int. Filter (s)}&{\bf Refinement (s)}&{\bf Total (s)} \\
		\hline
		
		&\multicolumn{6}{c|}{\bf 1000 T3 queries against T1}\\
		\hline
		{\bf RI}&{\bf 69.28\%}&28.60\%&{\bf 2.12\%}&0.52&0.10&0.64\\
		{\bf APRIL}&69.27\%&28.60\%&2.13\%&{\bf 0.06}&0.10&{\bf 0.18}\\
		\hline
		
		&\multicolumn{6}{c|}{\bf 1000 T3 queries against T2}\\
		\hline
		{\bf RI}&68.46\%&29.87\%&1.67\%&9.26&1.58&11.07\\
		{\bf APRIL}&68.46\%&29.87\%&1.67\%&{\bf 1.02}&1.58&{\bf 2.84}\\
		\hline
		
	\end{tabular}
\end{table*}
	
Next, we compare all methods in spatial {\em within} joins, where the objective is to find pairs $(r,s)$ such
that $r$ is within $s$ (see Section \ref{subsec:within}).
As Table \ref{tab:competitorswithin} shows, \methodname{} again achieves the
best performance, due to its extremely low filtering cost. 
\methodname{} is even faster than 5C+CH, because 5C+CH
performs two polygon-in-polygon tests which are slower compared to a
polygon intersection test.

\begin{table*}
	\centering
	\footnotesize
	\caption{Performance of filters (spatial
		within joins)}
	\label{tab:competitorswithin}
		\begin{tabular}{|l |@{~}c@{~}|@{~}c@{~}|@{~}c@{~}|@{~}c@{~}| @{~}c@{~}|@{~}c@{~}|}
			\hline
			&{\bf True hits}&{\bf True negatives}&{\bf Indecisive}&{\bf Int. Filter (s)}&{\bf Refinement (s)}&{\bf Total (s)} \\
			\hline
			&\multicolumn{6}{c|}{{\bf T2 $\bowtie$ T1 (Tiger water in landmark areas)}}\\
			\hline			
			{\bf None}&0.00\%&0.00\%&100.00\%&0.00&3.61&3.64\\
			{\bf 5C+CH}&0.00\%&34.71\%&65.29\%&0.10&1.33&1.46\\
			{\bf RA}&13.48\%&29.18\%&57.34\%&0.14&1.11&1.28\\
			{\bf RI}&{\bf 18.48\%}&{\bf 59.46\%}&{\bf 22.06\%}&0.20&{\bf 0.48}&0.71\\
			{\bf APRIL}&{\bf 18.48\%}&59.42\%&22.11\%&{\bf 0.05}&0.49&{\bf 0.58}\\
			\hline
			
			&\multicolumn{6}{c|}{{\bf T1 $\bowtie$ T3 (Tiger landmark in county areas)}}\\
			\hline
			{\bf None}&0.00\%&0.00\%&100.00\%&0.00&20.14&20.19\\
			{\bf 5C+CH}&0.00\%&20.72\%&79.28\%&0.37&14.02&14.44\\
			{\bf RA}&44.35\%&14.29\%&41.36\%&0.51&8.26&8.82\\
			{\bf RI}&{\bf 68.05\%}&{\bf 28.13\%}&{\bf 3.82\%}&1.56&{\bf 0.80}&2.41\\
			{\bf APRIL}&{\bf 68.05\%}&{\bf 28.13\%}&{\bf 3.82\%}&{\bf 0.21}&{\bf 0.80}&{\bf 1.06}\\
			\hline
			
			&\multicolumn{6}{c|}{{\bf T2 $\bowtie$ T3 (Tiger water in county areas)}}\\
			\hline
			{\bf None}&0.00\%&0.00\%&100.00\%&0.00&383.49&384.23\\
			{\bf 5C+CH}&0.00\%&22.17\%&77.83\%&7.70&274.54&282.98\\
			{\bf RA}&42.50\%&15.25\%&42.25\%&9.53&165.50&175.77\\
			{\bf RI}&{\bf 67.36\%}&{\bf 29.88\%}&{\bf 2.75\%}&27.08&{\bf 12.22}&40.04\\
			{\bf APRIL}&{\bf 67.36\%}&{\bf 29.88\%}&{\bf 2.75\%}&{\bf 3.47}&{\bf 12.22}&{\bf 16.43}\\
			\hline
		\end{tabular}
\end{table*}

Finally, we test the effectiveness of \methodname{} in
polygon-linestring joins, as described in Section
\ref{subsec:linestring}.
For this experiment, we join the polygon sets T1, T2, and T3 with
dataset T8 (from the same collection),
which contains 16.9M linestrings (roads in the United States), each having
20.4 vertices on average.
In this comparison, we do not include \oldmethod{} and RA, because
Strong cell types cannot be used to detect true hits.
Table \ref{tab:competitorslinestrings} compares 
\methodname{} with 5C+CH and the skipping of an intermediate filter
(None). 5C+CH only detects true negatives (in the case where the
5C+CH approximations do not intersect).
\methodname{} outperforms 5C+CH by at least three times in total join
time and by orders of magnitude in T3  $\bowtie$ T8, where it can
identify the great majority of join results as true hits.

\begin{table*}
	\centering
	\footnotesize
	\caption{Polygon-linestring spatial intersection joins.}
	\label{tab:competitorslinestrings}
	\begin{tabular}{|l |@{~}c@{~}|@{~}c@{~}|@{~}c@{~}|@{~}c@{~}| @{~}c@{~}|@{~}c@{~}|}
		\hline
		&{\bf True hits}&{\bf True negatives}&{\bf Indecisive}&{\bf Int. Filter (s)}&{\bf Refinement (s)}&{\bf Total (s)} \\
		\hline
		&\multicolumn{6}{c|}{{\bf T1 $\bowtie$ T8 (Tiger landmarks and roads)}}\\
		\hline
		{\bf None}&0.00\%&0.00\%&100.00\%&0.00&27.82&28.25\\
		{\bf 5C+CH}&0.00\%&45.24\%&54.76\%&1.07&15.99&17.49\\
		{\bf APRIL}&{\bf 12.70\%}&{\bf 55.01\%}&{\bf 32.29\%}&{\bf 0.93}&{\bf 3.82}&{\bf 5.18}\\
		\hline
		
		&\multicolumn{6}{c|}{{\bf T2 $\bowtie$ T8 (Tiger
				water areas and roads)}}\\
		\hline
		{\bf None}&0.00\%&0.00\%&100.00\%&0.00&238.91&241.59\\
		{\bf 5C+CH}&0.00\%&68.13\%&31.87\%&6.24&90.60&99.52\\
		{\bf APRIL}&{\bf 0.08\%}&{\bf 90.22\%}&{\bf 9.71\%}&{\bf 5.58}&{\bf 19.92}&{\bf 28.17}\\
		\hline
		
		&\multicolumn{6}{c|}{{\bf T3 $\bowtie$ T8 (Tiger
				county areas and roads)}}\\
		\hline
		{\bf None}&0.00\%&0.00\%&100.00\%&0.00&2546.48&2543.37\\
		{\bf 5C+CH}&0.00\%&22.79\%&77.21\%&16.21&1855.63&1878.73\\
		{\bf APRIL}&{\bf 66.25\%}&{\bf 30.77\%}&{\bf 2.98\%}&{\bf 25.64}&{\bf 58.23}&{\bf 90.77}\\
		\hline
		
	\end{tabular}
\end{table*}

%% file: relatedwork.tex
\section{Related Work}\label{sec:related}
Most previous works on spatial intersection joins \cite{JacoxS07} focus on the filter step of the join (denoted by MBR-join). They either exploit the pre-existing indexes \cite{BrinkhoffKS93,MamoulisP03} or partition the data on-the-fly and perform the join independently at each partition \cite{PatelD96,NobariTHKBA13,TsitsigkosBMT19}.
Each partition-to-partition MBR-join can be performed in memory with the help of plane-sweep \cite{BrinkhoffKS93,ArgePRSV98}.

\stitle{Intermediate filters}
To further reduce the candidate pairs that reach the refinement step,
conservative and/or progressive object approximations can be used for identifying false hits and/or true hits, respectively.  
Brinkhoff et al. \cite{BrinkhoffKSS94} suggested the use of the convex hull and the minimum bounding 5-corner convex polygon (5C) as conservative approximations and the maximum enclosing rectangle (MER) as a progressive approximation. MER is hard to compute and of questionable effectiveness \cite{ZimbraoS98}, hence, we did not include it in our comparison.
In follow-up work \cite{ZimbraoS98}, the object geometries are rasterized and modeled as grids, where each cell is colored based on its percentage of its coverage by the object. By re-scaling and aligning the grids of two candidate join objects, we can infer, in most cases, whether the objects are a join pair or a false hit. Indecisive pairs are forwarded to the refinement step.

\begin{figure}[htb]
	\centering
	\begin{tabular}{cc}
		\includegraphics[width=0.38\columnwidth]{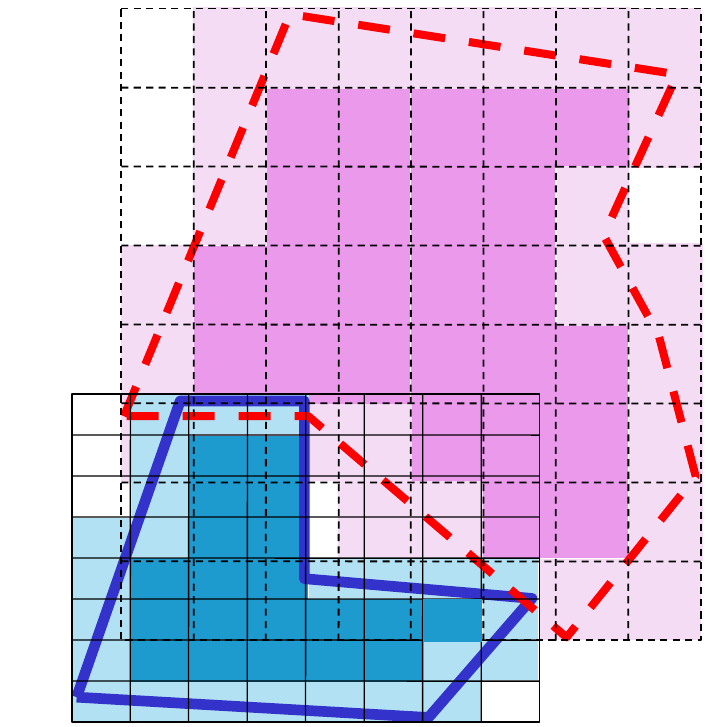}&
		\includegraphics[width=0.42\columnwidth]{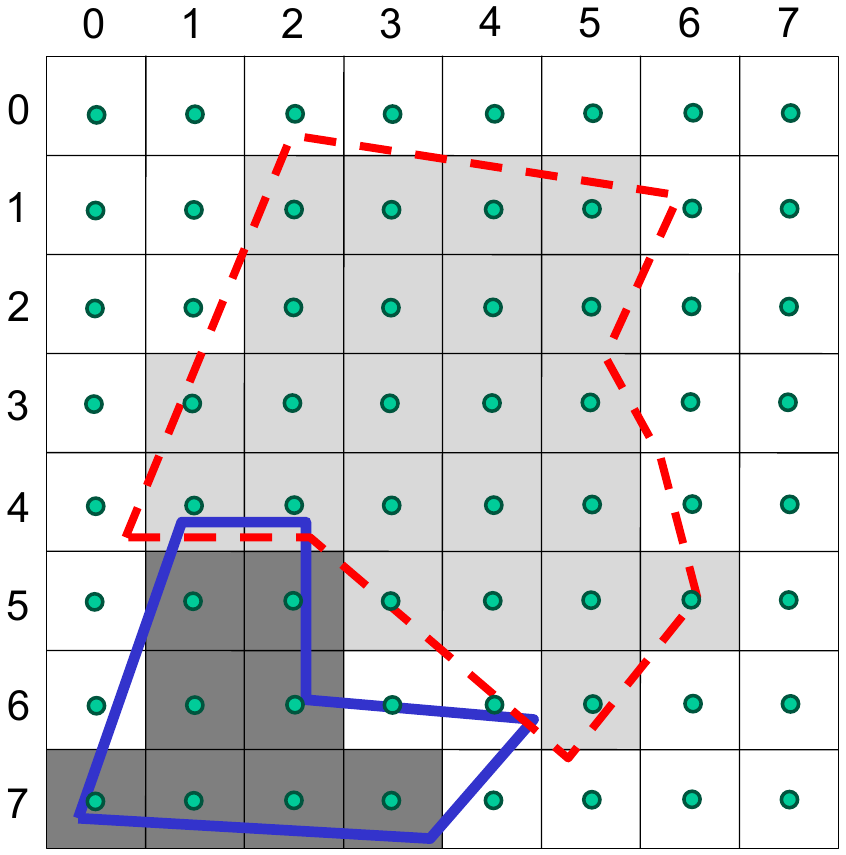}\\
		(a) \revj{IDEAL approximations}&
		(b) \revj{Raptor approximations} \\ 				
	\end{tabular}
	\caption{\revj{Examples of IDEAL and Raptor 
	object approximations.}}	
	\label{fig:raptor}
\end{figure}

\revj{
\stitle{Raster-based approaches for other queries}}
Hierarchical (quad-tree based) raster approximations based on a hierarchical grid have been used in the past \cite{DBLP:conf/sigmod/FangFNRS08} for window and distance queries. 
In addition, Teng et al. \cite{Teng21} propose IDEAL, a hybrid vector-raster polygonal approximation, targeting point-in-polygon queries and point-to-polygon distance queries.
\revj{
  The approximations in IDEAL are similar to those of APRIL in that they
capture information about Full or Partial coverage of each cell, but they
also have important differences that render IDEAL approximations not
appropriate for spatial intersection joins. Specifically, in IDEAL,
each polygon is approximated by its own (local) grid, defined by
splitting the object's MBR. Hence, the IDEAL grids of two different
objects are not necessarily aligned to each other and may have
different resolutions, as shown in Figure \ref{fig:raptor}(a).
Hence, IDEAL approximations are not appropriate
for intersection (or distance) joins because aligning the two different
grids of two polygons (having different positions, cell size, and
resolution) is hard and inference of polygon intersection from
the cell types if the cells are not perfectly aligned is not trivial.
Another difference between IDEAL and our work is that cells
in IDEAL are not grouped into intervals and interval joins are not
used as operations.}

\revj{RAPTOR \cite{SinglaE20,SinglaEDMS21,SinglaEDMS21b} joins a raster dataset (map of pixels, where each pixel is associated with values such as temperature) with a vector dataset
(e.g., set of polygons, linestrings, or points).
The objective of Raptor-Join is to identify, for each vector object $o$, the pixels that
are relevant to $o$
and associate
with $o$ the values of these pixels to the object (e.g., aggregate them).
For example, if the vector object $o$ is a polygon, the relevant pixels
to $o$ are those whose centroids are included in $o$. Hence, in Figure
\ref{fig:raptor}(b), the dashed polygon is relevant to the cells in
light-gray and the solid-border polygon is relevant to the cells
in dark-gray. To compute the Raptor-Join, in a pre-processing step,
all relevant cells to each object are identified and stored in a
tabular representation (called Flash Index),
with intervals of contiguous cells per row of
the raster matrix. For example, the  solid-border polygon is
represented by three tuples: $\{(5, [1,2]), (6, [1,2]), (7, [0,3])\}$,
implying that the polygon spans columns 1-2 in rows 5 and 6, and
columns 0-3 in row 7. This is reminiscent to our APRIL approximations,
where each object is represented by intervals of cells. However,
raster representations of vector objects used by RAPTOR have several important
differences to APRIL. First, for a cell to be included in a Raptor
approximation, the center of the cell should lie inside the
polygon, whereas in APRIL the cell should overlap with the polygon.
This means, for example, that cell (4,1) is not part of the
solid-border polygon approximation in Raptor, whereas it is part of
its APRIL approximation. Second, APRIL differentiates between Full and
Partial cells, whereas Raptor only has one type of cells. The most
important difference is that Raptor-join cannot be used for the
problem of spatial intersection joins that we study in this paper, as
it is possible that two polygons intersect but their Raptor
approximations share no common cell(s). In the example of Figure
\ref{fig:raptor}(b), the two polygons intersect each other in cells
(4,1), (4,2), and (6,4). Cells (4,1), (4,2) are included in the
approximation of the dashed-border polygon but not in the other one,
whereas cell (6,4) is included in neither of the two Raptor
approximations. Hence, Raptor-join, if applied for spatial
intersection joins, would mistakenly prune this pair of objects as
false positive.}

  \stitle{Speeding up the refinement step}
Checking whether two polygons overlap requires point-in-polygon tests and finding an intersection in the union of line segments that form both polygons \cite{BrinkhoffKSS94}. A point-in-polygon test bears a $O(n)$ cost, while the second problem can be solved in $O(n\log n)$ time \cite{ShamosH76}, where $n$ is the total number of edges in both polygons.
Given a pair of candidate objects,
Aghajarian et al. \cite{AghajarianPP16}
prune all line segments from the object geometries
that do not intersect their common MBR (CMBR) (i.e., the intersection area of their MBRs), before applying the refinement step.
This reduces the complexity of refinement, as a smaller number of segments need to be checked for intersection. In addition, if one object MBR is contained in the other, then the point-in-polygon test is applied before the segment intersection test.
Polysketch \cite{LiuYP19} decomposes each object to a set of tiles, i.e., small MBRs which include consecutive line segments of the object's geometry. Given two candidate objects, the refinement step is then applied only for the tile-pairs that overlap.
A similar idea (trapezoidal decomposition) was suggested by Brinkhoff et al. \cite{BrinkhoffKSS94} and alternative polygon decomposition approaches where suggested in \cite{BadawyA99}.
PSCMBR \cite{LiuP20a} combines Polysketch with the CMBR approach. Specifically, for the two candidate objects, the overlapping pairs of Polysketch tiles are found; for each such pair, the segments in the two tiles that do not overlap with the  CMBR of the tiles are pruned before refining the contents of the tiles. Polysketch and PSCMBR focus on finding the intersection points of two objects, hence, unlike our approach, they do not identify true hits. The CMBR approach \cite{AghajarianPP16}  is fully integrated in our implementation; still the refinement cost remains high.
Finally, the Clipped Bounding Box (CBB) \cite{SidlauskasCZA18}
is an enriched representation of the MBR that captures the dead (unused) space at MBR corners with
a few auxiliary points, providing the opportunity of refinement step avoidance in the case where object CBBs intersect only at their common dead-space areas.
CBBs can also be used by R-tree nodes to avoid their traversal if the query range overlaps only with their dead space.

\stitle{Approximate spatial joins}
The approximate representation of objects and approximate spatial query evaluation using  space-filling curves was first suggested by Orenstein \cite{Orenstein89}.
Recent work explores the use of raster approximations for the approximate evaluation of spatial joins and other operations \cite{KipfLPPAZDB0K20,ZacharatouKSPDM21,Kipf21}.
Our work is the first to approximate polygon rasterizations as intervals for {\em exact} spatial query evaluation. 

\stitle{Spatial joins on GPUs} The widespread availability of programmable GPUs has inspired several research efforts that leverage GPUs for spatial joins \cite{Sun03,AghajarianPP16,AghajarianP17,LiuYP19,LiuP20a}.
Sun et al. \cite{Sun03} accelerated the join refinement step by incorporating GPU rasterization as an intermediate filter.
This filter identifies \emph{only true negatives} using a low resolution, and has thus limited pruning effectiveness.  
Aghajarian et al. \cite{AghajarianPP16,AghajarianP17} proposed a GPU approach to process point-polygon and polygon-polygon joins for datasets that can be accommodated in GPU memory.
Liu et al. \cite{LiuYP19,LiuP20a} also proposed GPU-accelerated filters to reduce the number of refinements.  
These filters \cite{AghajarianPP16,AghajarianP17,LiuYP19,LiuP20a}, in contrast to \methodname{}, \emph{do not identify true hits}, but rather focus on finding the intersection points between a candidate pair.
Furthermore, the above approaches \cite{AghajarianPP16,AghajarianP17,LiuYP19,LiuP20a} do not involve rasterization and rely on CUDA, which is exclusive to NVIDIA GPUs.
A recent line of work \cite{ZacharatouDASF17,DoraiswamyF20,ZacharatouKSPDM21,DoraiswamyF22} proposes to use the GPU rasterization pipeline as an integral component of spatial query processing.
Doraiswamy et al. \cite{DoraiswamyF20,DoraiswamyF22} introduced a spatial data model and algebra that is designed to exploit modern GPUs. Their approach leverages a data representation called \emph{canvas}, which stores polygons as collections of pixels. 
The canvas includes a flag that differentiates between pixels that lie on the boundary of the polygon and those that are entirely covered by it.
Although current-generation GPUs can handle millions of polygons at fast frame rates, the evaluation of spatial queries is still dominated by other costs, such as triangulating polygons and performing I/Os \cite{DoraiswamyF22}.

\stitle{Scalability in spatial data management}
The emergence of cloud computing has led to many efforts to scale out spatial data management \cite{PandeyKNK18}. 
SJMP \cite{ZhangHLWX09} is an adaptation of the PBSM spatial join algorithm \cite{PatelD96} for MapReduce.
Other spatial data management systems that use MapReduce or Spark and handle spatial joins include Hadoop-GIS \cite{AjiWVLL0S13}, SpatialHadoop \cite{EldawyM15}, Magellan \cite{Magellan}, SpatialSpark \cite{YouZG15}, Simba \cite{XieL0LZG16}, and Apache Sedona \cite{YuZS19}.
\revj{These systems perform spatial partitioning of the data and distribution of the partitions to different machines, in order to improve scalability of spatial data management and query processing. Each partition is typically indexed by a data structure such as the R-tree or the quadtree, which facilitate spatial query evaluation.}
\revj{Regarding spatial intersection joins}, all the aforementioned systems focus only on the filter step \revj{and forward all candidate pairs directly to the refinement phase, which is implemented with the help of off-the-shelf libraries such as JTS\footnote{locationtech.github.io/jts/}. \methodname{} is orthogonal to the data partitioning approaches applied by these systems in the sense that its intermediate filter can follow any MBR filter or spatial index. The enhanced filtering of the added intermediate step reduces the amount of candidate pairs that must be geometrically refined, significantly improving the overall performance of spatial intersection join evaluation in them. Besides, as discussed in Section \ref{subsec:partitioning}, rasterization and APRIL object approximations can be applied independently to each one of the space-oriented partitions used by such systems.}

%% file: conclusion.tex
\vspace{-2mm}
\section{Conclusions}\label{sec:conclusions}
In this paper, we proposed a technique that represents
raster approximations of polygons as sets
of intervals, offering a fast and effective intermediate step between
the filter and the refinement steps of polygon intersection joins.
RI, the first version of our approach approximates each object as a
single list of intervals that include the raster cells that
intersect the object; together with each interval we store a bitstring
that encodes the classes of cells (Full, Strong, Weak) in the
interval.
APRIL is an enhanced version of our method that captures the cells
that are partially or fully covered by the object by two lists of
intervals and drops the space-consuming and burdensome bitstring.
APRIL's intermediate filter is different in that of RI in that it
performs a pipeline of three interval joins instead of a single
interval join paired with bitwise operations on the bitstrings.

As we have shown experimentally, compared to previous approaches
\cite{BrinkhoffKSS94,ZimbraoS98}, APRIL is (i) lightweight, as it
represents each polygon by two lists of integers that can be
effectively compressed; (ii) effective, as it typically filters the
majority of MBR-join pairs as true negatives or true positives; and
(iii) efficient to apply, as it only requires at most three linear
scans over the interval lists.
Specifically,
RI and APRIL offer at least 3x speedup in end-to-end spatial
intersection joins
compared to previous intermediate
filters (raster approximations \cite{ZimbraoS98}, 5C-CH \cite{BrinkhoffKSS94}). At the same time, the
space complexity of RI and APRIL is relatively low and the approximations can
easily be accommodated in main memory.
\revj{Compared to RI approximations, \methodname{}
  approximations are much cheaper to construct, occupy significantly
  less space, offer a much faster intermediate filter, and
  significantly improve the end-to-end cost of spatial intersection joins.}

\methodname{} is a general approximation for polygons that can also be
used in selection queries, within-joins and joins between polygons and
linestrings.
We propose a compression technique for \methodname{} and customizations that
trade space for filter effectiveness.
Finally, we propose an efficient construction technique for
\methodname{} approximations, which is orders of magnitude faster than 
rasterization-based techniques used for other filters.

In the future, we plan to investigate further the problem of interval join order optimization and explore the effectiveness of \methodname{} for 3D objects (e.g., polytopes).
\revj{We also aim to investigate integrating \methodname{}
  into a big distributed spatial
database management system, such as Apache Sedona as well as in
open-source spatial database systems such as PostGIS.}